\documentclass[prl,nobibnotes,floatfix,superscriptaddress,twocolumn]{revtex4-2}
\usepackage{amsfonts,amsmath,graphicx,epsfig,latexsym}
\usepackage{subfigure}
\usepackage{graphicx}
\usepackage{color}
\usepackage{natbib}
\usepackage{multirow}
\usepackage{latexsym,amssymb}
\usepackage{amsmath}
\usepackage{soul}
\usepackage{ulem}
\usepackage{array}
\usepackage{siunitx}
\newcolumntype{d}[1]{D{.}{.}{#1}}
\usepackage[linktocpage,bookmarksopen,bookmarksnumbered]{hyperref}
\usepackage[dvipsnames]{xcolor}

\usepackage{comment}
\usepackage{braket}

\usepackage{multirow}

\begin{document}

\title{Correlation Enhanced Electron-Phonon Coupling in FeSe/SrTiO$_3$ at a Magic Angle}




\author{Qiang Zou}
\thanks{These authors contributed equally to this work.}
\affiliation
{Department of Physics and Astronomy, West Virginia University, Morgantown, WV, USA}

\author{Antik Sihi}
\thanks{These authors contributed equally to this work.}
\affiliation
{Department of Physics and Astronomy, West Virginia University, Morgantown, WV, USA}
\thanks{These authors contributed equally to this work.}

\author{Basu Dev Oli}
\thanks{These authors contributed equally to this work.}
\affiliation
{Department of Physics and Astronomy, West Virginia University, Morgantown, WV, USA}
\thanks{These authors contributed equally to this work.}

\author{Mercè Roig}

\affiliation
{Department of Physics, University of Wisconsin--Milwaukee, Milwaukee, 53211, WI, USA}

\author{Daniel Agterberg}

\affiliation
{Department of Physics, University of Wisconsin--Milwaukee, Milwaukee, 53211, WI, USA}

\author{Michael Weinert}

\affiliation
{Department of Physics, University of Wisconsin--Milwaukee, Milwaukee, 53211, WI, USA}

\author{Lian Li}
\email{Contact author: lian.li@mail.wvu.edu}
\affiliation
{Department of Physics and Astronomy, West Virginia University, Morgantown, WV, USA}

\author{Subhasish Mandal}
\email{Contact author: subhasish.mandal@mail.wvu.edu}
\affiliation
{Department of Physics and Astronomy, West Virginia University, Morgantown, WV, USA}



\begin{abstract}
{\footnotesize

While a predictive theory for unconventional superconductivity in Fe-based superconductors remains elusive, an extensively debated aspect is the interaction between phonons and strongly correlated electrons, and its potential role in the pairing mechanism. Here, through the combination of first principles dynamical mean field theory calculations and epitaxial growth of the single-layer FeX (X=Se, S, Te) on SrTiO$_3$ (STO)(001) substrate, which facilitates the controlled distortion of the FeX$_4$ tetrahedron, we demonstrate an unique superconducting dome where the superconducting gap peaks at a `magic' angle of the FeX$_4$ tetrahedron and the electron-phonon coupling (EPC) for the A$_{1g}$ mode is maximized for the FeSe film. Our findings uncover a significant role of electronic correlations in strengthening Cooper pairing in unconventional superconductors by enhancing EPC.

}
\end{abstract}



\maketitle



{\it Introduction.} While a well-developed predictive theory through electron-phonon coupling (EPC) exists for conventional superconductors (SC), the mechanism responsible for electron pairing in unconventional high-temperature superconductors remains one of the most profound open questions in condensed matter physics. Both theoretical predictions and experimental observations have revealed significant lattice coupling with the strongly interacting $d$-electrons in superconducting cuprates~\cite{cuprate-EPC1,cuprate-EPC2,cuprate-EPC3,cuprate-EPC4,cuprate-EPC5,cuprate-EPC6,cuprate-EPC7,cuprate-EPC8}—and more recently in the nickelates~\cite{EPC-nickel3,EPC-nickel1,EPC-nickel2}—suggesting that EPC may play a significant role in unconventional superconductivity~\cite{FeSC-SM1,EPC_FeSe,FeSe-STO1,FeSe-STO2,FeSe-STO3,EPC-nature,Coh2015,EPC-Ba122}. Although there is compelling evidence across multiple families of unconventional SCs~\cite{cuprate-EPC1,cuprate-EPC2,cuprate-EPC3,cuprate-EPC4,cuprate-EPC5,cuprate-EPC6,cuprate-EPC7,EPC-nickel3,EPC-nickel1,EPC-nickel2,FeSC-SM1,EPC_FeSe,FeSe-STO1,FeSe-STO2,FeSe-STO3,EPC-nature,Coh2015,EPC-Ba122,phonon-KH}, achieving simultaneous and systematic tuning of both the superconducting transition temperature ($T_c$) and the strength of EPC through atomic-scale structural modifications remains a significant bottleneck.

Iron-based superconductors provide a unique platform to explore this puzzle. These materials are layered systems comprising planes of edge-sharing FeAs(Se)$_4$ tetrahedra, interleaved with charge reservoir or spacer layers. For bulk Fe-pnictides, a phenomenological correlation has been observed between $T_c$ and the As–Fe–As bond angle, with optimal superconductivity near a critical tetrahedral geometry~\cite{angle-tc-2,angle-tc}. In these systems, unconventional superconductivity is believed to emerge in close proximity to a quantum critical point associated with a competing symmetry-broken nematic phase~\cite{Keimer2015,nematic1,nematic2,nematic3}. 

A particularly intriguing case is single-layer (SL) FeSe grown on SrTiO$_3$(001) (FeSe/STO), which exhibits a $T_c$ nearly an order of magnitude higher than that of bulk FeSe~\cite{FeSe-STO1,FeSe-STO2,FeSe-STO3,EPC-nature} and coincides with a suppression of nematicity~\cite{SLFeSe_nematic_suppression}. Importantly, interface-specific phenomena, such as enhanced EPC with interfacial phonon modes, have been proposed as key contributors to the dramatic $T_c$ enhancement in this system~\cite{FeSe-STO1,FeSe-STO2,FeSe-STO3,EPC-nature}. Despite intense experimental and theoretical focus, the precise microscopic origin of the enhanced superconductivity in FeSe/STO remains unresolved. There are still open questions, such as: what is the relative contribution of EPC of the FeSe film and that of the interface and their connection to electron correlations to the enhanced $T_c$? Can EPC be quantitatively tuned by modifying the geometry of the FeSe layer or the underlying substrate? Addressing these challenges is critical for advancing our fundamental understanding of unconventional superconductivity and for designing new quantum materials with enhanced superconductivity.

Here, by integrating {\it ab initio} computations based on Density functional theory (DFT)+ embedded Dynamical Mean Field Theory (eDMFT), heterostructure design, epitaxial growth, scanning tunneling spectroscopy/microscopy (STS/M), and Angle-resolved Photoemission Spectroscopy (ARPES) measurements, we demonstrate that the key factor driving the much-enhanced superconductivity in SL FeSe/STO(001) is the enhancement of EPC within the FeSe film, driven by strong electronic correlations. In contrast to previous interpretations attributing the enhancement to interfacial EPC~\cite{FeSe-STO1,FeSe-STO2,FeSe-STO3,EPC-nature}, our results highlight the dominant role of correlation-enhanced EPC within the FeSe film itself. To controllably alter the tetrahedral geometry in the SL FeX/STO, we epitaxially grow SL FeX (X=Se, S, Te), and their alloys on TiO$_2$- and SrO-terminated STO. Epitaxial growth of SL FeX alloy films on the same STO substrate thus allows the direct comparison of their tetrahedral geometry in SL FeX/STO films. Similar to the UV-light induced structural change~\cite{UV-Lian}, this offers an effective route to manipulate the X--Fe--X angle. 

\begin{figure}[t]
\centering
\includegraphics[width=0.5\textwidth]{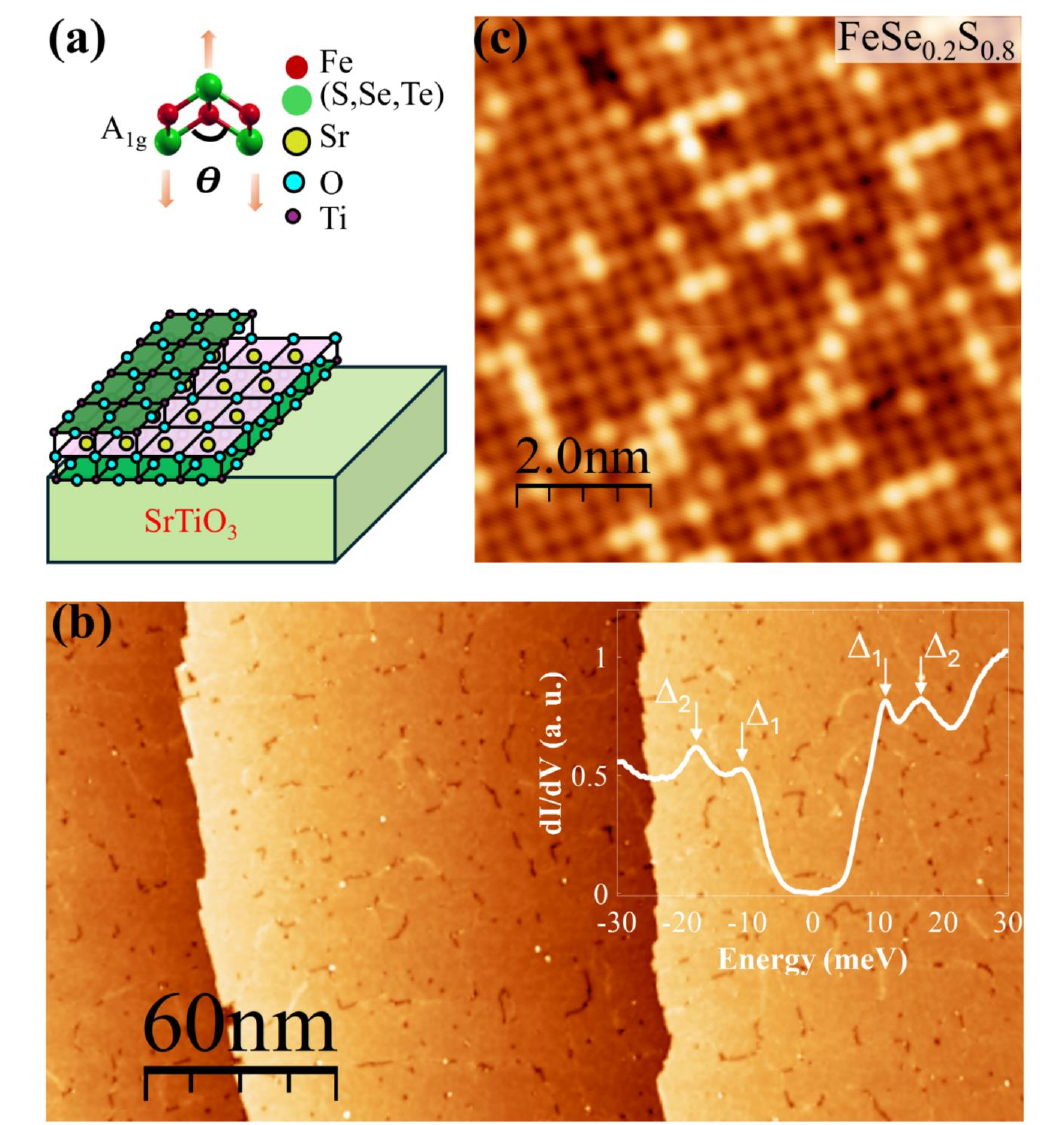}
\caption{(a) Ball-and-stick model of the single-layer FeX/STO with the A$_{1g}$ mode marked by the arrows. STM images of single layer (b) FeSe/STO and (c) FeSe$_{0.2}$S$_{0.8}$/STO taken at 4.3 K. Inset shows dI/dV spectra for the single-layer FeSe/STO.}
\label{fig1}
\end{figure}

We demonstrate a distinct superconducting dome driven by the X--Fe--X bond angle, where charge transfer from the STO substrate, electron correlations, and phonon coupling act cooperatively. Variations in STO termination (TiO$_2$ vs.\ SrO), oxygen vacancies, and Te/S alloying modulate both the bond angle and the degree of charge transfer. We show that as the alloy compositions and substrate terminations are changed from FeSe$_x$Te$_{1-x}$/STO to FeSe$_y$S$_{1-y}$/STO, the measured superconducting gap ($\Delta$) peaks for a `magic' tetrahedral bond angle of $\sim$107\textdegree. Computed strength of electron correlations is found to substantially decrease with S alloying, which agrees well with the experiment. Most significantly, our eDMFT-computed deformation potential associated with the A$_{1g}$ phonon mode (Fig.~\ref{fig1}\textbf{a}), which is proportional to the EPC, exhibits a dome similar to $\Delta$ as a function of X--Fe--X angle. These findings establish a direct link between atomic-scale structural modifications, modulation of EPC strength, and $T_c$, highlighting the potential for geometrically engineered control over superconductivity in FeSe/STO systems.

{\it Results:} Experimentally, we grow high-quality single-layer FeX films on SrTiO$_3$(001) (STO) substrates using molecular beam epitaxy (MBE). Fig.~\ref{fig1}(b) shows the surface morphology of the films, which conforms to the step-terrace structure of the STO substrate. While the composition can be controlled precisely during growth by adjusting the k-cell temperature supplying the X, the growth process also inherently introduces chemical and electronic inhomogeneities at the atomic scale. For the intermediate alloy compositions, we have developed a machine-learning solution to analyze the microscopic STS/M data and identify patterns and correlations that may not be immediately apparent through visual inspection alone~\cite{Lian-3,Lian-4}.

\begin{figure}[t]
\centering
\includegraphics[width=0.50\textwidth]{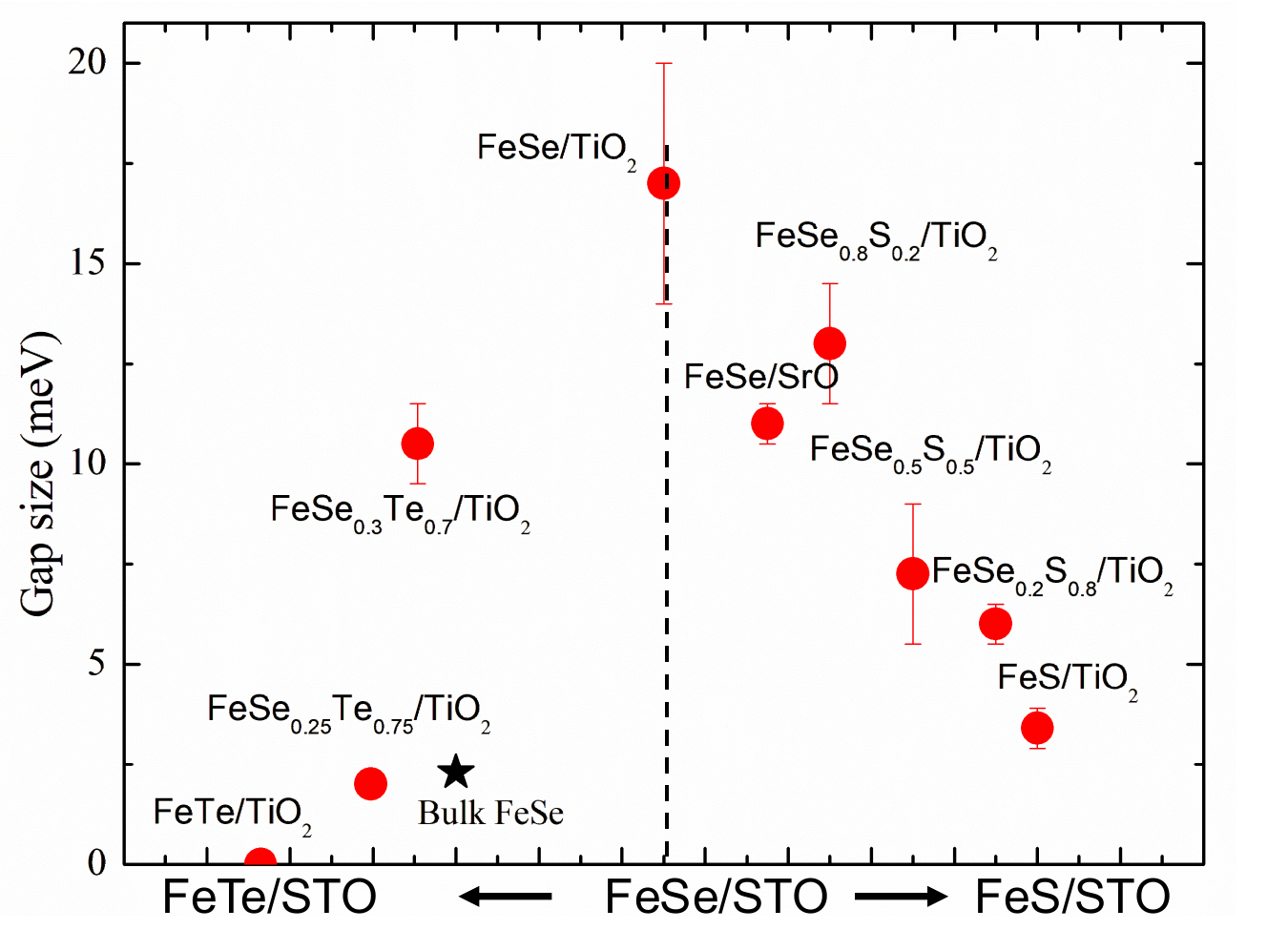}
\caption{Summary of the superconducting gap for the single layer FeX/STO showing a superconducting dome as a function of doping Te and S. The gap for bulk FeSe is obtained from Ref.~\cite{doi:10.1126/science.1202226} and indicated as a star.}\label{fig3}

\end{figure}

After careful calibration of the Se, S, and Te ratio, we check the evolution of the $\Delta$ using STS measurements at temperatures between 0.35 and 4.3 K, as shown in Fig.~\ref{fig3} and Supplementary Fig.~2. For SL FeSe, the dI/dV spectrum is fully gapped (U-shaped) around the Fermi level. For higher S (Te) concentrations, the dI/dV spectra are V-shaped, with a smaller superconducting coherence peak at 5 $\pm$ 1 meV. For FeS, the dI/dV spectra remained gaped with a coherence peak at 3 $\pm$ 1 meV. In contrast, while the dI/dV spectra are V-shaped, no coherence peak can be discerned, indicating that FeTe is not superconducting. The results are summarized in Fig.~\ref{fig3}, where a dome peaked at FeSe/TiO$_2$ is seen. Note that the $T_c$ is $\sim$ 60 K for FeSe/TiO2, determined by our own {\it in situ} temperature-dependent ARPES measurements~\cite{UV-Lian}. The evolution of the Fermi surface (FS) for SL FeX films is taken at 80 K, as shown in Supplementary Figs. 3 and 4. 

 

\begin{figure}[t]
\centering
\includegraphics[width=0.50\textwidth]{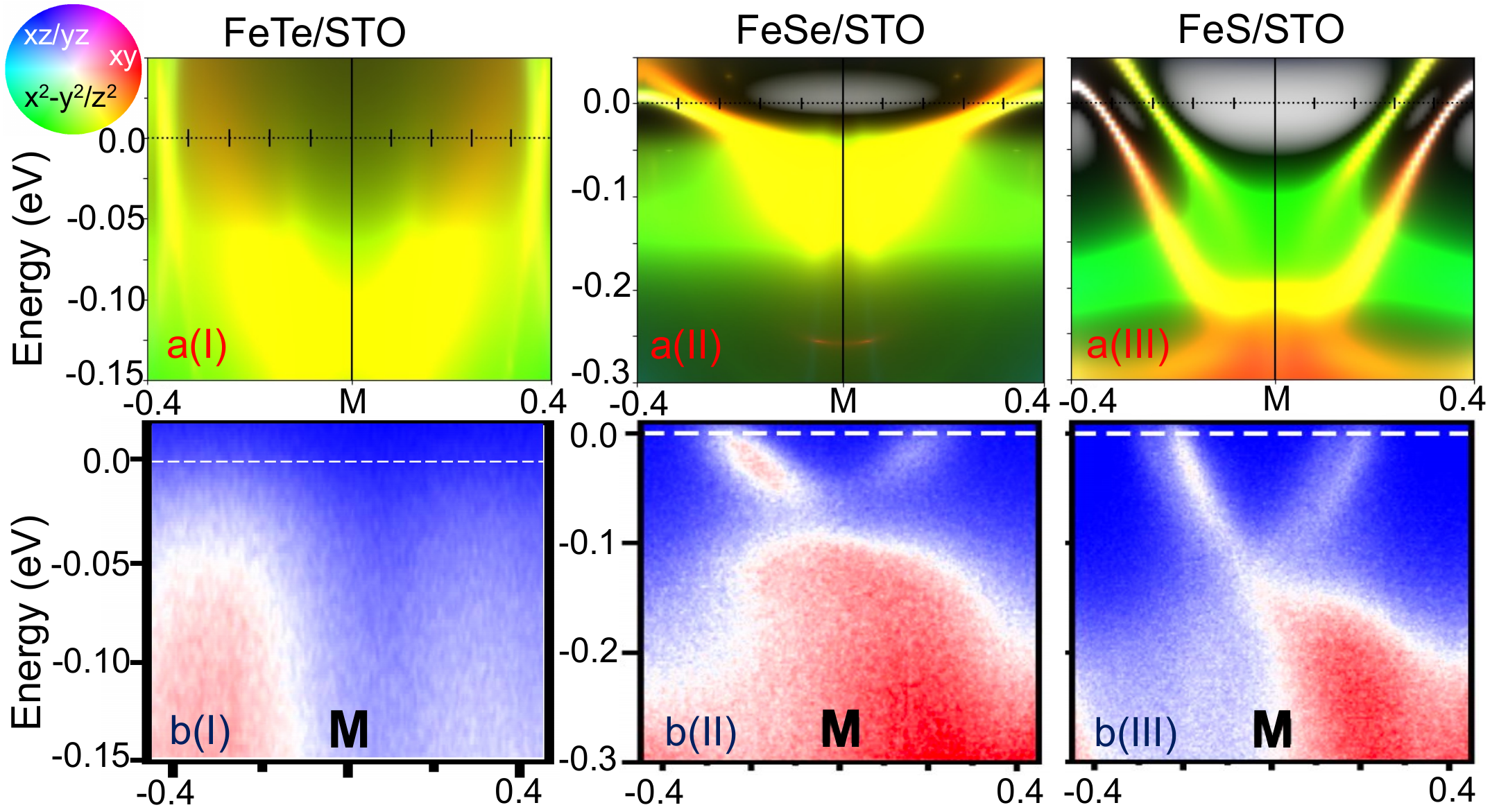}
\caption{Comparison of spectral functions obtained from eDMFT computations [a(I)--a(III)] and ARPES measurements [b(I)--b(III)] as a function of S/Te-alloying in single layer FeX/STO.}\label{fig-SF}

\end{figure}

\begin{figure*}[t]
\centering
\includegraphics[width=1\textwidth]{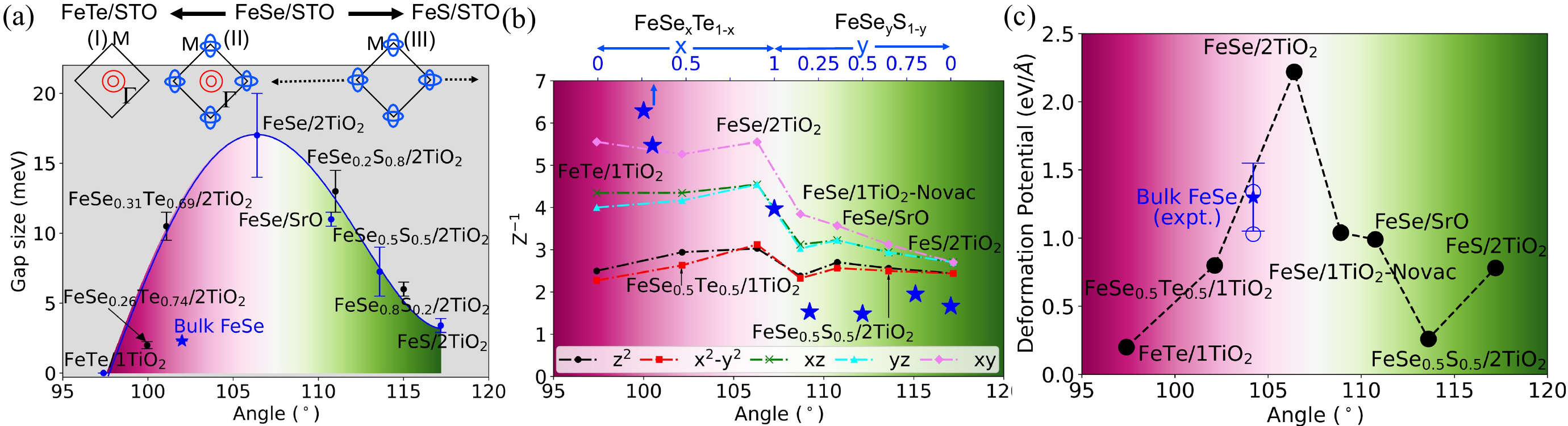}
\caption{
\textbf{The combined role of electron correlation and EPC from eDMFT calculations.} (a)  Experimental superconducting gap size ($\Delta$) as measured in the STS showing a superconducting dome as a function of eDMFT computed X--Fe--X angle. The gap for bulk FeSe is obtained from Ref.~\cite{doi:10.1126/science.1202226} and indicated as a star. eDMFT computed angles are marked with blue dots, while TEM and extrapolated angles are marked with black dots. (inset) Schematic representation of the evolution of the experimental Fermi surface of FeSe$_{1-x}$Te (S)$_x$/SrTiO$_3$ interfaces as a function of Te (S) concentration. (b) Computed mass-enhancement for all five Fe-{\it d} orbitals and as a function of X--Fe--X angle in various heterostructures; the blue stars in (b) represents average experimental mass enhancement of the $M$ pocket as a function of Te and S concentration (top axis). (c) Deformation potential ($\mathcal{D}$) of the A$_{1g}$ phonon mode computed using eDMFT, evaluated along the experimentally measured Fermi surface, showing a dome similar to panel (a). $\mathcal{D}$ for bulk FeSe are obtained from Ref.~\cite{EPC_FeSe}, where blue star with error bar represents the value from the direct measurement of $\mathcal{D}$ and two blue open circles represent eDMFT computed average and maximum $\mathcal{D}$ as obtained from Ref.~\cite{FeSC-SM1,EPC_FeSe}. The pink (green) shaded area is for Te (S) doping, while the white area is for the FeSe/STO. }\label{fig4}
\end{figure*}

To understand the mechanism behind the  enhanced  superconductivity in SL FeX/STO from first principles eDMFT method, we describe our computational work in Figs.~\ref{fig-SF} and~\ref{fig4}. We optimize the atom positions in the unit cell using eDMFT for various FeX/STO heterostructures with 50\% O-vacancy on either 1-TiO$_2$ or 2-TiO$_2$ terminated STO, namely FeTe/1TiO$_2$, FeTe$_{0.5}$Se$_{0.5}$/1TiO$_2$, FeSe/2TiO$_2$, FeSe$_{0.5}$S$_{0.5}$/2TiO$_2$, and FeS/2TiO$_2$. To understand the role of O-vacancy and substrate terminations, we also investigate FeSe/STO without O-vacancy (FeSe/1TiO$_2$-Novac) and FeSe/SrO heterostructures. Detailed atomic structures are described in the Supplementary Fig.~10, Supplementary Tables 1 and 2. The atomic positions of various heterostructures including the substrate atoms are included in the optimization process that incorporates the effect of the electron’s entropy while computing forces on atoms~\cite{Forces}. The X--Fe--X bond angles (X = S, Se, Te) optimized using eDMFT in the FeX/STO heterostructures differ significantly from those obtained via standard DFT (see Supplementary Information (SI)) and show good agreement with previously reported experimental values~\cite{charles-sciadv,substrate-Mandal}.

It is particularly insightful to examine how the spectral functions (SF) evolve upon alloying SL FeSe/STO with S and Te. In Fig.~\ref{fig-SF} and in End Matter (Fig.~\ref{fig4_em}), we illustrate this evolution and compare  results from ARPES and eDMFT calculations. SFs from theory and experiment show an excellent agreement and capture the evolution of both electron and hole pockets driven by S/Te alloying, which is primarily due to charge transfer from the substrate to the FeX film (Fig.~\ref{fig-SF} and Fig.~\ref{fig4_em}). In addition, eDMFT accurately reproduces the trends in electronic correlation strength across the alloy series, consistent with ARPES measurements. For example, increasing Te concentration leads to a clear enhancement of correlations, as reflected in both the electron pockets (Fig.~\ref{fig-SF}) and the hole pockets (Figs.~\ref{fig4_em}d(I)--d(V)), compared to the sharper, more coherent bands observed in FeS/STO.


Next, in Fig.~\ref{fig4}\textbf{a}, we show the dramatic evolution of the FS in a schematic diagram, summarizing the ARPES results in Supplementary Fig. 3 as a function of concentration in FeTe/STO $\rightarrow$ FeSe/STO $\rightarrow$ FeS/STO.  The Te-alloyed region has FS containing only central ($\Gamma$) pockets (Fig.~\ref{fig4}\textbf{a(I)}) where the S-alloyed region has only electron pockets at the zone corners (Fig.~\ref{fig4}\textbf{a(III)}). For the intermediate Te-rich side (Fig.~\ref{fig4}\textbf{a(II)}), both the electron and hole pockets are present as described in Supplementary Fig.~3\textbf{e}. To explain the phenomenological correlation of $T_c$ and X--Fe--X angle, the eDMFT-computed angles are then plotted against the measured $\Delta$ in Fig.~\ref{fig4}\textbf{a}. Since some of the heterostructures have asymmetric chalcogen heights, we take the average of the top and bottom layers of X--Fe--X. These are marked with blue dots in Fig.~\ref{fig4}\textbf{a}. For some alloy compositions, where a direct optimization from eDMFT computations was not possible due to a very large supercell, we estimate the angles using Vegard's law relating the lattice constant to composition for binary alloys. These are marked with black dots in Fig.~\ref{fig4}\textbf{a}. Overall, $\Delta$ as a function of X--Fe--X angle shows an interesting dome-like behavior; with increasing angle, $\Delta$ first increases and reaches a maximum and then decreases. 


Next, we describe orbital-resolved mass enhancements (compared to DFT mass) to investigate the effect of electron correlation as a function of the X--Fe--X angles (Fig.~\ref{fig4}\textbf{b}) directly from the continuous-time Monte Carlo sampling. First, across all doping levels, the mass enhancement is found to be strongly orbital dependent, with the $xy$ orbital exhibiting the highest degree of correlation (it is labeled according to the folded one Fe-atom equivalent unit cell). This pronounced orbital selectivity is a hallmark of Hund's metal behavior~\cite{Haule-njp09,yin2011kinetic,Mandal:2014,Roser2015,luca-1}. The degree of orbital differentiation reaches its maximum in the FeTe/2TiO$_2$, where all orbitals exhibit strong correlations. This effect gradually weakens with increasing S incorporation and is least pronounced in the FeS/STO. The electron correlations remains high near the magic angle in the FeSe/2TiO$_2$, but decreases when the substrate is replaced with SrO-terminated STO or 1TiO$_2$-terminated STO without O-vacancies. A similar trend in electron correlation with O-vacancy was found in FeSe/1TiO$_2$~\cite{mandal-PRL}. With further increase in the X--Fe--X bond angle, particularly in the S-doped regime, the correlation strength drops sharply. The variation in correlation strength with S (Te) alloying in FeSe/STO—decreasing with increasing S content and increasing with Te concentration—is in strong agreement with our ARPES measurements as discussed above and as seen by the blue stars in Fig.~\ref{fig4}\textbf{b}, which represents the average experimental mass enhancement of the $M$ pocket as a function of Te and S concentration (top axis).  It is interesting to note that the SL FeTe/STO, with or without oxygen vacancies, exhibits a weaker correlations compared to its bulk counterpart, where strong correlation effects have been observed experimentally and confirmed by recent eDMFT and ARPES studies~\cite{Chris-arxiv}. The effect of possible antiferromagnetic ordering in SL FeTe/STO is not considered here. Overall, this highlights the dual role of substrate terminations in controlling the electronic structure of SL FeX/STO: by modulating the carrier level, as noticed by moving the Fermi energy (Supplementary Fig.~11) and by tuning the electron correlations through changes in the X--Fe--X bond angle. 



 Earlier eDMFT-computations  have predicted that electron correlations can significantly enhance the strength of EPC in bulk FeSe~\cite{FeSC-SM1}. Such prediction has been verified by a recent experiment employing ultrafast techniques which enabled direct determination of the EPC strength by combining both time-resolved ARPES (trARPES) and time-resolved X-ray diffraction (trXRD)~\cite{EPC_FeSe}. To systematically understand the critical behavior of the superconductivity as a function of X--Fe--X angle, we investigate the vibrational properties of various SL FeX/STO heterostructures for the A$_{1g}$ mode, which is associated with the X--Fe--X angle. 
 
 First, we find that the computed A$_{1g}$ frequency for FeSe/2TiO$_2$ is 22.6 meV, which is in strong agreement with both Raman scattering~\cite{PhysRevB.87.144508} and ultrafast experiments~\cite{EPC_FeSe}. Next, we compute the deformation potential ($\mathcal{D}$) for A$_{1g}$ mode in SL FeX/STO using eDMFT. The EPC is proportional to the square of $\mathcal{D}$, which is the ratio of the displacement of energy with respect to the displacement of the ion (details in SI). From the optimized structures, the $\mathcal{D}$ is computed (Supplementary Fig.~12\textbf{a}) for the X--Fe--X layer for the electron pockets and hole pockets on a dense k-mesh. The average $\mathcal{D}$ over electron pockets (blue) and hole pockets (red) are plotted in Supplementary Fig.~12\textbf{a}. It is noteworthy that, in the free-standing X--Fe--X layer, both electron and hole bands are present near the Fermi level, in contrast to experimental ARPES results. This is a consequence of not including the substrate atoms explicitly in the computations of $\mathcal{D}$. Consequently, the value of $\mathcal{D}$ corresponding to the experimental FS (black circle in Fig.~\ref{fig4}\textbf{c}) is taken to be either the average $\mathcal{D}$ of the electron pockets or that of the hole pockets, depending on the FS features observed in ARPES for a given heterostructure. For example, in FeTe/STO, where ARPES reveals only hole pockets, the experimental $\mathcal{D}$ corresponds to that of the hole pockets. Conversely, for heterostructures such as FeSe/2TiO$_2$, where ARPES shows only electron pockets at the Brillouin zone corner, $\mathcal{D}$ for the experimental FS is taken as the average $\mathcal{D}$ of the electron pockets.  We aligned the equilibrium X–Fe–X angle of the free-standing SL with the optimized angle obtained from the corresponding heterostructure. We then computed $\mathcal{D}$ only for the X--Fe--X layers giving a finite displacement to the X atoms. This way we capture the substrate-induced doping effect and incorporate the interface-driven modifications to the bond angle directly on the vibrational properties of the SL FeX films without explicitly including the substrate atoms in our calculations. This approach enables us to circumvent the significant computational challenges associated with explicitly including the substrate atoms for $\mathcal{D}$ calculation, based on the assumption that their effect on the EPC for the A$_{1g}$ mode on the experimental FS is minimal. Fig.~\ref{fig4}\textbf{c} shows that the computed $\mathcal{D}$ for the experimental FS behaves similar to the $\Delta$ as a function of X--Fe--X angle (Fig. ~\ref{fig4}\textbf{a}), which first increases and then decreases with increasing angle. The dome-like behavior in $\mathcal{D}$ is observed for the entire alloy range, except in FeSe$_{0.5}$S$_{0.5}$, which is an outlier and has less $\mathcal{D}$ than FeS/STO despite the angle being smaller. This behavior may be attributed to the suppression of the A$_{1g}$ phonon mode within this concentration range, which is consistent with observations in bulk FeSe$_{1-x}$S$_x$, where the A$_{1g}$ mode vanishes at x=0.23 and re-emerges at x=0.69 with significantly higher phonon energy~\cite{FeSeS}. Notably, the trend in $\mathcal{D}$ as a function of the X--Fe--X angle aligns with the direct measurement of $\mathcal{D}$ for bulk FeSe, determined through combined trARPES and trXRD~\cite{EPC_FeSe}, and shown as the blue star in Fig.~\ref{fig4}\textbf{c}.

To compare the effect of the same A$_{1g}$ phonon's interaction with electrons treated in the mean-field level, we perform the same computations for determining  the angles and the $\mathcal{D}$ using standard DFT and find that the $\mathcal{D}$ in DFT does not show any dome-like behavior with Te and S incorporation (Supplementary Fig.~12). The detailed comparison is described in Supplemental Tables 1 and 2. The enhanced EPC observed in DMFT, relative to DFT, and the unique angle dependence originates from the many-body interactions among correlated electrons in FeSe/STO--effects that are not accounted for within the single-particle framework of conventional DFT~\cite{FeSC-SM1,EPC_Georges1}. Further, since FeSe exhibits strong characteristics of a Hund’s metal~\cite{Haule-njp09,yin2011kinetic,Mandal:2014,luca-1}, this enhancement goes beyond commonly used approaches that mimic the magnetic fluctuations through static magnetic moments.  

{\it Conclusions:} Previous eDMFT calculations for the bulk FeSe indicate that the A$_{1g}$ mode is the dominant contributor (roughly 40\%) to the total EPC. Since the A$_{1g}$ phonon mode belongs to zero momentum transfer (\textbf{q}=0) phonon and there are only electron pockets in the FS in the optimal 2-TiO$_2$ FeSe/STO, this forward electron-phonon scattering with \textbf{q}=0 (or any small-\textbf{q}) phonon can enhance superconductivity in the same way irrespective of its electronic or phonon origin~\cite{Lee:2015,FeSe-STO3,FeSe-STO-natcom,arxiv-EPC}. Considering the pockets at the $M$-point, the effective electron-electron interaction mediated by the $A_{1g}$ phonons in the singlet channel, can indeed enhance superconductivity both in the $s$-wave and nodeless $d$-wave channels (see SI for details). The combination of spin fluctuations and the small \textbf{q} EPC, as we found here in the SL FeSe/STO, can lead to the most enhanced superconductivity in 2-TiO$_2$ FeSe/STO.  
The estimated EPC in eDMFT for this A$_{1g}$ mode in optimal 2-TiO$_2$ FeSe/STO is found to be $\sim$ 6 (7) times higher compared to its bulk (DFT) value, which can yield a T$_c$ of $\sim$ 24 (33)K, following the Allen-Dynes formula with the phonon frequency of $\sim$ 213K (see SI). Together with the interfacial EPC found in various experiments~\cite{EPC-nature,FeSC-EPC1}, the giant EPC for the optimal X--Fe--X angle can explain the much enhanced superconductivity in 2-TiO$_2$ FeSe/STO. Alternatively, if all phonon modes are uniformly enhanced by optimal electronic correlations, EPC alone may be sufficient to account for the high $T_c$ observed in SL FeSe/STO. Thus, this work offers new insights into the pairing mechanism in SL FeSe/STO from a phononic standpoint, highlighting the possibility of incorporating electronic correlation effects into the canonical BCS framework. Moreover, our findings pave the way for tailoring superconductivity in SL Fe-SCs and other families of high-$T_c$ materials, such as nickelates, through precise control of atomic-scale geometry.


\textbf{Acknowledgments} S.M. thanks Kristjan Haule for stimulating discussions. S.M. acknowledges the support from the Air Force Office of Scientific Research by the Department of Defense under Award No. FA9550-23-1-0498 of the DEPSCoR program. S.M. and A.S. benefited from the Frontera supercomputer at the Texas Advanced Computing Center (TACC) at The University of Texas at Austin, which is supported by National Science Foundation Grant No. OAC-1818253. A.S. also acknowledges the support from NSF OAC-2311558. L.L. and M.W. were supported by the U.S. Department of Energy, Office of Basic Energy Sciences, Award No. DE-SC0021393 and DE-SC0017632. D.~F.~A. was supporetd  by the U.S. Department of Energy, Office of Basic Energy Sciences, Division of Materials Sciences and Engineering under Award No. DE-SC0021971. Work at UWM  was also supported by a grant from the Simons Foundation (SFI-MPS-NFS-00006741-02, D.F.A and M.R.).
S.M. and L.L. conceived and designed the project. S.M and A.S. performed all the calculations and developed the code for the EPC calculation with eDMFT. Q.Z. and B.D.O grew the thin films and performed the ARPES and STM/S experiment. M.R., D.A., M.W. performed the modeling of superconductivity. S.M. and L.L. analyzed and interpreted the experimental and computational data. All authors participated in discussions and in writing of the manuscript. 

\bibliography{sn-bibliography}

\begin{thebibliography}{48}%
\makeatletter
\providecommand \@ifxundefined [1]{%
 \@ifx{#1\undefined}
}%
\providecommand \@ifnum [1]{%
 \ifnum #1\expandafter \@firstoftwo
 \else \expandafter \@secondoftwo
 \fi
}%
\providecommand \@ifx [1]{%
 \ifx #1\expandafter \@firstoftwo
 \else \expandafter \@secondoftwo
 \fi
}%
\providecommand \natexlab [1]{#1}%
\providecommand \enquote  [1]{``#1''}%
\providecommand \bibnamefont  [1]{#1}%
\providecommand \bibfnamefont [1]{#1}%
\providecommand \citenamefont [1]{#1}%
\providecommand \href@noop [0]{\@secondoftwo}%
\providecommand \href [0]{\begingroup \@sanitize@url \@href}%
\providecommand \@href[1]{\@@startlink{#1}\@@href}%
\providecommand \@@href[1]{\endgroup#1\@@endlink}%
\providecommand \@sanitize@url [0]{\catcode `\\12\catcode `\$12\catcode `\&12\catcode `\#12\catcode `\^12\catcode `\_12\catcode `\%12\relax}%
\providecommand \@@startlink[1]{}%
\providecommand \@@endlink[0]{}%
\providecommand \url  [0]{\begingroup\@sanitize@url \@url }%
\providecommand \@url [1]{\endgroup\@href {#1}{\urlprefix }}%
\providecommand \urlprefix  [0]{URL }%
\providecommand \Eprint [0]{\href }%
\providecommand \doibase [0]{https://doi.org/}%
\providecommand \selectlanguage [0]{\@gobble}%
\providecommand \bibinfo  [0]{\@secondoftwo}%
\providecommand \bibfield  [0]{\@secondoftwo}%
\providecommand \translation [1]{[#1]}%
\providecommand \BibitemOpen [0]{}%
\providecommand \bibitemStop [0]{}%
\providecommand \bibitemNoStop [0]{.\EOS\space}%
\providecommand \EOS [0]{\spacefactor3000\relax}%
\providecommand \BibitemShut  [1]{\csname bibitem#1\endcsname}%
\let\auto@bib@innerbib\@empty
\bibitem [{\citenamefont {Lee}\ \emph {et~al.}(2006)\citenamefont {Lee}, \citenamefont {Fujita}, \citenamefont {McElroy}, \citenamefont {Slezak}, \citenamefont {Wang}, \citenamefont {Aiura}, \citenamefont {Bando}, \citenamefont {Ishikado}, \citenamefont {Masui}, \citenamefont {Zhu}, \citenamefont {Balatsky}, \citenamefont {Eisaki}, \citenamefont {Uchida},\ and\ \citenamefont {Davis}}]{cuprate-EPC1}%
  \BibitemOpen
  \bibfield  {author} {\bibinfo {author} {\bibfnamefont {J.}~\bibnamefont {Lee}}, \bibinfo {author} {\bibfnamefont {K.}~\bibnamefont {Fujita}}, \bibinfo {author} {\bibfnamefont {K.}~\bibnamefont {McElroy}}, \bibinfo {author} {\bibfnamefont {J.~A.}\ \bibnamefont {Slezak}}, \bibinfo {author} {\bibfnamefont {M.}~\bibnamefont {Wang}}, \bibinfo {author} {\bibfnamefont {Y.}~\bibnamefont {Aiura}}, \bibinfo {author} {\bibfnamefont {H.}~\bibnamefont {Bando}}, \bibinfo {author} {\bibfnamefont {M.}~\bibnamefont {Ishikado}}, \bibinfo {author} {\bibfnamefont {T.}~\bibnamefont {Masui}}, \bibinfo {author} {\bibfnamefont {J.~X.}\ \bibnamefont {Zhu}}, \bibinfo {author} {\bibfnamefont {A.~V.}\ \bibnamefont {Balatsky}}, \bibinfo {author} {\bibfnamefont {H.}~\bibnamefont {Eisaki}}, \bibinfo {author} {\bibfnamefont {S.}~\bibnamefont {Uchida}},\ and\ \bibinfo {author} {\bibfnamefont {J.~C.}\ \bibnamefont {Davis}},\ }\bibfield  {title} {\bibinfo {title} {Interplay of electron--lattice interactions and superconductivity in
  bi2sr2cacu2o8+$\delta$},\ }\href {https://doi.org/10.1038/nature04973} {\bibfield  {journal} {\bibinfo  {journal} {Nature}\ }\textbf {\bibinfo {volume} {442}},\ \bibinfo {pages} {546} (\bibinfo {year} {2006})}\BibitemShut {NoStop}%
\bibitem [{\citenamefont {Krakauer}\ \emph {et~al.}(1993)\citenamefont {Krakauer}, \citenamefont {Pickett},\ and\ \citenamefont {Cohen}}]{cuprate-EPC2}%
  \BibitemOpen
  \bibfield  {author} {\bibinfo {author} {\bibfnamefont {H.}~\bibnamefont {Krakauer}}, \bibinfo {author} {\bibfnamefont {W.~E.}\ \bibnamefont {Pickett}},\ and\ \bibinfo {author} {\bibfnamefont {R.~E.}\ \bibnamefont {Cohen}},\ }\bibfield  {title} {\bibinfo {title} {Large calculated electron-phonon interactions in ${\mathrm{la}}_{2\mathrm{\ensuremath{-}}\mathit{x}}$${\mathit{m}}_{\mathit{x}}$${\mathrm{cuo}}_{4}$},\ }\href {https://doi.org/10.1103/PhysRevB.47.1002} {\bibfield  {journal} {\bibinfo  {journal} {Phys. Rev. B}\ }\textbf {\bibinfo {volume} {47}},\ \bibinfo {pages} {1002} (\bibinfo {year} {1993})}\BibitemShut {NoStop}%
\bibitem [{\citenamefont {Devereaux}\ \emph {et~al.}(2004)\citenamefont {Devereaux}, \citenamefont {Cuk}, \citenamefont {Shen},\ and\ \citenamefont {Nagaosa}}]{cuprate-EPC3}%
  \BibitemOpen
  \bibfield  {author} {\bibinfo {author} {\bibfnamefont {T.~P.}\ \bibnamefont {Devereaux}}, \bibinfo {author} {\bibfnamefont {T.}~\bibnamefont {Cuk}}, \bibinfo {author} {\bibfnamefont {Z.-X.}\ \bibnamefont {Shen}},\ and\ \bibinfo {author} {\bibfnamefont {N.}~\bibnamefont {Nagaosa}},\ }\bibfield  {title} {\bibinfo {title} {Anisotropic electron-phonon interaction in the cuprates},\ }\href {https://doi.org/10.1103/PhysRevLett.93.117004} {\bibfield  {journal} {\bibinfo  {journal} {Phys. Rev. Lett.}\ }\textbf {\bibinfo {volume} {93}},\ \bibinfo {pages} {117004} (\bibinfo {year} {2004})}\BibitemShut {NoStop}%
\bibitem [{\citenamefont {Lanzara}\ \emph {et~al.}(2001)\citenamefont {Lanzara}, \citenamefont {Bogdanov}, \citenamefont {Zhou}, \citenamefont {Kellar}, \citenamefont {Feng}, \citenamefont {Lu}, \citenamefont {Yoshida}, \citenamefont {Eisaki}, \citenamefont {Fujimori}, \citenamefont {Kishio}, \citenamefont {Shimoyama}, \citenamefont {Noda}, \citenamefont {Uchida}, \citenamefont {Hussain},\ and\ \citenamefont {Shen}}]{cuprate-EPC4}%
  \BibitemOpen
  \bibfield  {author} {\bibinfo {author} {\bibfnamefont {A.}~\bibnamefont {Lanzara}}, \bibinfo {author} {\bibfnamefont {P.~V.}\ \bibnamefont {Bogdanov}}, \bibinfo {author} {\bibfnamefont {X.~J.}\ \bibnamefont {Zhou}}, \bibinfo {author} {\bibfnamefont {S.~A.}\ \bibnamefont {Kellar}}, \bibinfo {author} {\bibfnamefont {D.~L.}\ \bibnamefont {Feng}}, \bibinfo {author} {\bibfnamefont {E.~D.}\ \bibnamefont {Lu}}, \bibinfo {author} {\bibfnamefont {T.}~\bibnamefont {Yoshida}}, \bibinfo {author} {\bibfnamefont {H.}~\bibnamefont {Eisaki}}, \bibinfo {author} {\bibfnamefont {A.}~\bibnamefont {Fujimori}}, \bibinfo {author} {\bibfnamefont {K.}~\bibnamefont {Kishio}}, \bibinfo {author} {\bibfnamefont {J.~I.}\ \bibnamefont {Shimoyama}}, \bibinfo {author} {\bibfnamefont {T.}~\bibnamefont {Noda}}, \bibinfo {author} {\bibfnamefont {S.}~\bibnamefont {Uchida}}, \bibinfo {author} {\bibfnamefont {Z.}~\bibnamefont {Hussain}},\ and\ \bibinfo {author} {\bibfnamefont {Z.~X.}\ \bibnamefont {Shen}},\ }\bibfield  {title} {\bibinfo {title}
  {Evidence for ubiquitous strong electron--phonon coupling in high-temperature superconductors},\ }\href {https://doi.org/10.1038/35087518} {\bibfield  {journal} {\bibinfo  {journal} {Nature}\ }\textbf {\bibinfo {volume} {412}},\ \bibinfo {pages} {510} (\bibinfo {year} {2001})}\BibitemShut {NoStop}%
\bibitem [{\citenamefont {Iwasawa}\ \emph {et~al.}(2008)\citenamefont {Iwasawa}, \citenamefont {Douglas}, \citenamefont {Sato}, \citenamefont {Masui}, \citenamefont {Yoshida}, \citenamefont {Sun}, \citenamefont {Eisaki}, \citenamefont {Bando}, \citenamefont {Ino}, \citenamefont {Arita}, \citenamefont {Shimada}, \citenamefont {Namatame}, \citenamefont {Taniguchi}, \citenamefont {Tajima}, \citenamefont {Uchida}, \citenamefont {Saitoh}, \citenamefont {Dessau},\ and\ \citenamefont {Aiura}}]{cuprate-EPC5}%
  \BibitemOpen
  \bibfield  {author} {\bibinfo {author} {\bibfnamefont {H.}~\bibnamefont {Iwasawa}}, \bibinfo {author} {\bibfnamefont {J.~F.}\ \bibnamefont {Douglas}}, \bibinfo {author} {\bibfnamefont {K.}~\bibnamefont {Sato}}, \bibinfo {author} {\bibfnamefont {T.}~\bibnamefont {Masui}}, \bibinfo {author} {\bibfnamefont {Y.}~\bibnamefont {Yoshida}}, \bibinfo {author} {\bibfnamefont {Z.}~\bibnamefont {Sun}}, \bibinfo {author} {\bibfnamefont {H.}~\bibnamefont {Eisaki}}, \bibinfo {author} {\bibfnamefont {H.}~\bibnamefont {Bando}}, \bibinfo {author} {\bibfnamefont {A.}~\bibnamefont {Ino}}, \bibinfo {author} {\bibfnamefont {M.}~\bibnamefont {Arita}}, \bibinfo {author} {\bibfnamefont {K.}~\bibnamefont {Shimada}}, \bibinfo {author} {\bibfnamefont {H.}~\bibnamefont {Namatame}}, \bibinfo {author} {\bibfnamefont {M.}~\bibnamefont {Taniguchi}}, \bibinfo {author} {\bibfnamefont {S.}~\bibnamefont {Tajima}}, \bibinfo {author} {\bibfnamefont {S.}~\bibnamefont {Uchida}}, \bibinfo {author} {\bibfnamefont {T.}~\bibnamefont {Saitoh}},
  \bibinfo {author} {\bibfnamefont {D.~S.}\ \bibnamefont {Dessau}},\ and\ \bibinfo {author} {\bibfnamefont {Y.}~\bibnamefont {Aiura}},\ }\bibfield  {title} {\bibinfo {title} {Isotopic fingerprint of electron-phonon coupling in high-${T}_{c}$ cuprates},\ }\href {https://doi.org/10.1103/PhysRevLett.101.157005} {\bibfield  {journal} {\bibinfo  {journal} {Phys. Rev. Lett.}\ }\textbf {\bibinfo {volume} {101}},\ \bibinfo {pages} {157005} (\bibinfo {year} {2008})}\BibitemShut {NoStop}%
\bibitem [{\citenamefont {Reznik}\ \emph {et~al.}(2006)\citenamefont {Reznik}, \citenamefont {Pintschovius}, \citenamefont {Ito}, \citenamefont {Iikubo}, \citenamefont {Sato}, \citenamefont {Goka}, \citenamefont {Fujita}, \citenamefont {Yamada}, \citenamefont {Gu},\ and\ \citenamefont {Tranquada}}]{cuprate-EPC6}%
  \BibitemOpen
  \bibfield  {author} {\bibinfo {author} {\bibfnamefont {D.}~\bibnamefont {Reznik}}, \bibinfo {author} {\bibfnamefont {L.}~\bibnamefont {Pintschovius}}, \bibinfo {author} {\bibfnamefont {M.}~\bibnamefont {Ito}}, \bibinfo {author} {\bibfnamefont {S.}~\bibnamefont {Iikubo}}, \bibinfo {author} {\bibfnamefont {M.}~\bibnamefont {Sato}}, \bibinfo {author} {\bibfnamefont {H.}~\bibnamefont {Goka}}, \bibinfo {author} {\bibfnamefont {M.}~\bibnamefont {Fujita}}, \bibinfo {author} {\bibfnamefont {K.}~\bibnamefont {Yamada}}, \bibinfo {author} {\bibfnamefont {G.~D.}\ \bibnamefont {Gu}},\ and\ \bibinfo {author} {\bibfnamefont {J.~M.}\ \bibnamefont {Tranquada}},\ }\bibfield  {title} {\bibinfo {title} {Electron--phonon coupling reflecting dynamic charge inhomogeneity in copper oxide superconductors},\ }\href {https://doi.org/10.1038/nature04704} {\bibfield  {journal} {\bibinfo  {journal} {Nature}\ }\textbf {\bibinfo {volume} {440}},\ \bibinfo {pages} {1170} (\bibinfo {year} {2006})}\BibitemShut {NoStop}%
\bibitem [{\citenamefont {Braicovich}\ \emph {et~al.}(2020)\citenamefont {Braicovich}, \citenamefont {Rossi}, \citenamefont {Fumagalli}, \citenamefont {Peng}, \citenamefont {Wang}, \citenamefont {Arpaia}, \citenamefont {Betto}, \citenamefont {De~Luca}, \citenamefont {Di~Castro}, \citenamefont {Kummer}, \citenamefont {Moretti~Sala}, \citenamefont {Pagetti}, \citenamefont {Balestrino}, \citenamefont {Brookes}, \citenamefont {Salluzzo}, \citenamefont {Johnston}, \citenamefont {van~den Brink},\ and\ \citenamefont {Ghiringhelli}}]{cuprate-EPC7}%
  \BibitemOpen
  \bibfield  {author} {\bibinfo {author} {\bibfnamefont {L.}~\bibnamefont {Braicovich}}, \bibinfo {author} {\bibfnamefont {M.}~\bibnamefont {Rossi}}, \bibinfo {author} {\bibfnamefont {R.}~\bibnamefont {Fumagalli}}, \bibinfo {author} {\bibfnamefont {Y.}~\bibnamefont {Peng}}, \bibinfo {author} {\bibfnamefont {Y.}~\bibnamefont {Wang}}, \bibinfo {author} {\bibfnamefont {R.}~\bibnamefont {Arpaia}}, \bibinfo {author} {\bibfnamefont {D.}~\bibnamefont {Betto}}, \bibinfo {author} {\bibfnamefont {G.~M.}\ \bibnamefont {De~Luca}}, \bibinfo {author} {\bibfnamefont {D.}~\bibnamefont {Di~Castro}}, \bibinfo {author} {\bibfnamefont {K.}~\bibnamefont {Kummer}}, \bibinfo {author} {\bibfnamefont {M.}~\bibnamefont {Moretti~Sala}}, \bibinfo {author} {\bibfnamefont {M.}~\bibnamefont {Pagetti}}, \bibinfo {author} {\bibfnamefont {G.}~\bibnamefont {Balestrino}}, \bibinfo {author} {\bibfnamefont {N.~B.}\ \bibnamefont {Brookes}}, \bibinfo {author} {\bibfnamefont {M.}~\bibnamefont {Salluzzo}}, \bibinfo {author} {\bibfnamefont
  {S.}~\bibnamefont {Johnston}}, \bibinfo {author} {\bibfnamefont {J.}~\bibnamefont {van~den Brink}},\ and\ \bibinfo {author} {\bibfnamefont {G.}~\bibnamefont {Ghiringhelli}},\ }\bibfield  {title} {\bibinfo {title} {Determining the electron-phonon coupling in superconducting cuprates by resonant inelastic x-ray scattering: Methods and results on ${\mathrm{nd}}_{1+x}{\mathrm{ba}}_{2\ensuremath{-}x}{\mathrm{cu}}_{3}{\mathrm{o}}_{7\ensuremath{-}\ensuremath{\delta}}$},\ }\href {https://doi.org/10.1103/PhysRevResearch.2.023231} {\bibfield  {journal} {\bibinfo  {journal} {Phys. Rev. Res.}\ }\textbf {\bibinfo {volume} {2}},\ \bibinfo {pages} {023231} (\bibinfo {year} {2020})}\BibitemShut {NoStop}%
\bibitem [{\citenamefont {Wang}\ \emph {et~al.}(2021)\citenamefont {Wang}, \citenamefont {von Arx}, \citenamefont {Horio}, \citenamefont {Mukkattukavil}, \citenamefont {K{\"u}spert}, \citenamefont {Sassa}, \citenamefont {Schmitt}, \citenamefont {Nag}, \citenamefont {Pyon}, \citenamefont {Takayama}, \citenamefont {Takagi}, \citenamefont {Garcia-Fernandez}, \citenamefont {Zhou},\ and\ \citenamefont {Chang}}]{cuprate-EPC8}%
  \BibitemOpen
  \bibfield  {author} {\bibinfo {author} {\bibfnamefont {Q.}~\bibnamefont {Wang}}, \bibinfo {author} {\bibfnamefont {K.}~\bibnamefont {von Arx}}, \bibinfo {author} {\bibfnamefont {M.}~\bibnamefont {Horio}}, \bibinfo {author} {\bibfnamefont {D.~J.}\ \bibnamefont {Mukkattukavil}}, \bibinfo {author} {\bibfnamefont {J.}~\bibnamefont {K{\"u}spert}}, \bibinfo {author} {\bibfnamefont {Y.}~\bibnamefont {Sassa}}, \bibinfo {author} {\bibfnamefont {T.}~\bibnamefont {Schmitt}}, \bibinfo {author} {\bibfnamefont {A.}~\bibnamefont {Nag}}, \bibinfo {author} {\bibfnamefont {S.}~\bibnamefont {Pyon}}, \bibinfo {author} {\bibfnamefont {T.}~\bibnamefont {Takayama}}, \bibinfo {author} {\bibfnamefont {H.}~\bibnamefont {Takagi}}, \bibinfo {author} {\bibfnamefont {M.}~\bibnamefont {Garcia-Fernandez}}, \bibinfo {author} {\bibfnamefont {K.-J.}\ \bibnamefont {Zhou}},\ and\ \bibinfo {author} {\bibfnamefont {J.}~\bibnamefont {Chang}},\ }\bibfield  {title} {\bibinfo {title} {Charge order lock-in by electron-phonon coupling in
  ${La}_{1.675}$${Eu}_{0.2}$${Sr}_{0.125}$${CuO}_{4}$},\ }\href {https://doi.org/10.1126/sciadv.abg7394} {\bibfield  {journal} {\bibinfo  {journal} {Science Advances}\ }\textbf {\bibinfo {volume} {7}},\ \bibinfo {pages} {eabg7394} (\bibinfo {year} {2021})},\ \Eprint {https://arxiv.org/abs/https://www.science.org/doi/pdf/10.1126/sciadv.abg7394} {https://www.science.org/doi/pdf/10.1126/sciadv.abg7394} \BibitemShut {NoStop}%
\bibitem [{\citenamefont {Meier}\ \emph {et~al.}(2024)\citenamefont {Meier}, \citenamefont {de~Vaulx}, \citenamefont {Bernardini}, \citenamefont {Botana}, \citenamefont {Blase}, \citenamefont {Olevano},\ and\ \citenamefont {Cano}}]{EPC-nickel3}%
  \BibitemOpen
  \bibfield  {author} {\bibinfo {author} {\bibfnamefont {Q.~N.}\ \bibnamefont {Meier}}, \bibinfo {author} {\bibfnamefont {J.~B.}\ \bibnamefont {de~Vaulx}}, \bibinfo {author} {\bibfnamefont {F.}~\bibnamefont {Bernardini}}, \bibinfo {author} {\bibfnamefont {A.~S.}\ \bibnamefont {Botana}}, \bibinfo {author} {\bibfnamefont {X.}~\bibnamefont {Blase}}, \bibinfo {author} {\bibfnamefont {V.}~\bibnamefont {Olevano}},\ and\ \bibinfo {author} {\bibfnamefont {A.}~\bibnamefont {Cano}},\ }\bibfield  {title} {\bibinfo {title} {Preempted phonon-mediated superconductivity in the infinite-layer nickelates},\ }\href {https://doi.org/10.1103/PhysRevB.109.184505} {\bibfield  {journal} {\bibinfo  {journal} {Phys. Rev. B}\ }\textbf {\bibinfo {volume} {109}},\ \bibinfo {pages} {184505} (\bibinfo {year} {2024})}\BibitemShut {NoStop}%
\bibitem [{\citenamefont {Li}\ and\ \citenamefont {Louie}(2024)}]{EPC-nickel1}%
  \BibitemOpen
  \bibfield  {author} {\bibinfo {author} {\bibfnamefont {Z.}~\bibnamefont {Li}}\ and\ \bibinfo {author} {\bibfnamefont {S.~G.}\ \bibnamefont {Louie}},\ }\bibfield  {title} {\bibinfo {title} {Two-gap superconductivity and the decisive role of rare-earth $d$ electrons in infinite-layer nickelates},\ }\href {https://doi.org/10.1103/PhysRevLett.133.126401} {\bibfield  {journal} {\bibinfo  {journal} {Phys. Rev. Lett.}\ }\textbf {\bibinfo {volume} {133}},\ \bibinfo {pages} {126401} (\bibinfo {year} {2024})}\BibitemShut {NoStop}%
\bibitem [{\citenamefont {Zhan}\ \emph {et~al.}(2025)\citenamefont {Zhan}, \citenamefont {Gu}, \citenamefont {Wu},\ and\ \citenamefont {Hu}}]{EPC-nickel2}%
  \BibitemOpen
  \bibfield  {author} {\bibinfo {author} {\bibfnamefont {J.}~\bibnamefont {Zhan}}, \bibinfo {author} {\bibfnamefont {Y.}~\bibnamefont {Gu}}, \bibinfo {author} {\bibfnamefont {X.}~\bibnamefont {Wu}},\ and\ \bibinfo {author} {\bibfnamefont {J.}~\bibnamefont {Hu}},\ }\bibfield  {title} {\bibinfo {title} {Cooperation between electron-phonon coupling and electronic interaction in bilayer nickelates ${\mathrm{la}}_{3}{\mathrm{ni}}_{2}{\mathrm{o}}_{7}$},\ }\href {https://doi.org/10.1103/PhysRevLett.134.136002} {\bibfield  {journal} {\bibinfo  {journal} {Phys. Rev. Lett.}\ }\textbf {\bibinfo {volume} {134}},\ \bibinfo {pages} {136002} (\bibinfo {year} {2025})}\BibitemShut {NoStop}%
\bibitem [{\citenamefont {Mandal}\ \emph {et~al.}(2014{\natexlab{a}})\citenamefont {Mandal}, \citenamefont {Cohen},\ and\ \citenamefont {Haule}}]{FeSC-SM1}%
  \BibitemOpen
  \bibfield  {author} {\bibinfo {author} {\bibfnamefont {S.}~\bibnamefont {Mandal}}, \bibinfo {author} {\bibfnamefont {R.~E.}\ \bibnamefont {Cohen}},\ and\ \bibinfo {author} {\bibfnamefont {K.}~\bibnamefont {Haule}},\ }\bibfield  {title} {\bibinfo {title} {Strong pressure-dependent electron-phonon coupling in fese},\ }\href {https://doi.org/10.1103/PhysRevB.89.220502} {\bibfield  {journal} {\bibinfo  {journal} {Phys. Rev. B}\ }\textbf {\bibinfo {volume} {89}},\ \bibinfo {pages} {220502(R)} (\bibinfo {year} {2014}{\natexlab{a}})}\BibitemShut {NoStop}%
\bibitem [{\citenamefont {Gerber}\ \emph {et~al.}(2017)\citenamefont {Gerber}, \citenamefont {Yang}, \citenamefont {Zhu}, \citenamefont {Soifer}, \citenamefont {Sobota}, \citenamefont {Rebec}, \citenamefont {Lee}, \citenamefont {Jia}, \citenamefont {Moritz}, \citenamefont {Jia}, \citenamefont {Gauthier}, \citenamefont {Li}, \citenamefont {Leuenberger}, \citenamefont {Zhang}, \citenamefont {Chaix}, \citenamefont {Li}, \citenamefont {Jang}, \citenamefont {Lee}, \citenamefont {Yi}, \citenamefont {Dakovski}, \citenamefont {Song}, \citenamefont {Glownia}, \citenamefont {Nelson}, \citenamefont {Kim}, \citenamefont {Chuang}, \citenamefont {Hussain}, \citenamefont {Moore}, \citenamefont {Devereaux}, \citenamefont {Lee}, \citenamefont {Kirchmann},\ and\ \citenamefont {Shen}}]{EPC_FeSe}%
  \BibitemOpen
  \bibfield  {author} {\bibinfo {author} {\bibfnamefont {S.}~\bibnamefont {Gerber}}, \bibinfo {author} {\bibfnamefont {S.-L.}\ \bibnamefont {Yang}}, \bibinfo {author} {\bibfnamefont {D.}~\bibnamefont {Zhu}}, \bibinfo {author} {\bibfnamefont {H.}~\bibnamefont {Soifer}}, \bibinfo {author} {\bibfnamefont {J.~A.}\ \bibnamefont {Sobota}}, \bibinfo {author} {\bibfnamefont {S.}~\bibnamefont {Rebec}}, \bibinfo {author} {\bibfnamefont {J.~J.}\ \bibnamefont {Lee}}, \bibinfo {author} {\bibfnamefont {T.}~\bibnamefont {Jia}}, \bibinfo {author} {\bibfnamefont {B.}~\bibnamefont {Moritz}}, \bibinfo {author} {\bibfnamefont {C.}~\bibnamefont {Jia}}, \bibinfo {author} {\bibfnamefont {A.}~\bibnamefont {Gauthier}}, \bibinfo {author} {\bibfnamefont {Y.}~\bibnamefont {Li}}, \bibinfo {author} {\bibfnamefont {D.}~\bibnamefont {Leuenberger}}, \bibinfo {author} {\bibfnamefont {Y.}~\bibnamefont {Zhang}}, \bibinfo {author} {\bibfnamefont {L.}~\bibnamefont {Chaix}}, \bibinfo {author} {\bibfnamefont {W.}~\bibnamefont {Li}}, \bibinfo
  {author} {\bibfnamefont {H.}~\bibnamefont {Jang}}, \bibinfo {author} {\bibfnamefont {J.-S.}\ \bibnamefont {Lee}}, \bibinfo {author} {\bibfnamefont {M.}~\bibnamefont {Yi}}, \bibinfo {author} {\bibfnamefont {G.~L.}\ \bibnamefont {Dakovski}}, \bibinfo {author} {\bibfnamefont {S.}~\bibnamefont {Song}}, \bibinfo {author} {\bibfnamefont {J.~M.}\ \bibnamefont {Glownia}}, \bibinfo {author} {\bibfnamefont {S.}~\bibnamefont {Nelson}}, \bibinfo {author} {\bibfnamefont {K.~W.}\ \bibnamefont {Kim}}, \bibinfo {author} {\bibfnamefont {Y.-D.}\ \bibnamefont {Chuang}}, \bibinfo {author} {\bibfnamefont {Z.}~\bibnamefont {Hussain}}, \bibinfo {author} {\bibfnamefont {R.~G.}\ \bibnamefont {Moore}}, \bibinfo {author} {\bibfnamefont {T.~P.}\ \bibnamefont {Devereaux}}, \bibinfo {author} {\bibfnamefont {W.-S.}\ \bibnamefont {Lee}}, \bibinfo {author} {\bibfnamefont {P.~S.}\ \bibnamefont {Kirchmann}},\ and\ \bibinfo {author} {\bibfnamefont {Z.-X.}\ \bibnamefont {Shen}},\ }\bibfield  {title} {\bibinfo {title} {Femtosecond
  electron-phonon lock-in by photoemission and x-ray free-electron laser},\ }\href {https://doi.org/10.1126/science.aak9946} {\bibfield  {journal} {\bibinfo  {journal} {Science}\ }\textbf {\bibinfo {volume} {357}},\ \bibinfo {pages} {71} (\bibinfo {year} {2017})},\ \Eprint {https://arxiv.org/abs/https://www.science.org/doi/pdf/10.1126/science.aak9946} {https://www.science.org/doi/pdf/10.1126/science.aak9946} \BibitemShut {NoStop}%
\bibitem [{\citenamefont {Ge}\ \emph {et~al.}(2015)\citenamefont {Ge}, \citenamefont {Liu}, \citenamefont {Liu}, \citenamefont {Gao}, \citenamefont {Qian}, \citenamefont {Xue}, \citenamefont {Liu},\ and\ \citenamefont {Jia}}]{FeSe-STO1}%
  \BibitemOpen
  \bibfield  {author} {\bibinfo {author} {\bibfnamefont {J.-F.}\ \bibnamefont {Ge}}, \bibinfo {author} {\bibfnamefont {Z.-L.}\ \bibnamefont {Liu}}, \bibinfo {author} {\bibfnamefont {C.}~\bibnamefont {Liu}}, \bibinfo {author} {\bibfnamefont {C.-L.}\ \bibnamefont {Gao}}, \bibinfo {author} {\bibfnamefont {D.}~\bibnamefont {Qian}}, \bibinfo {author} {\bibfnamefont {Q.-K.}\ \bibnamefont {Xue}}, \bibinfo {author} {\bibfnamefont {Y.}~\bibnamefont {Liu}},\ and\ \bibinfo {author} {\bibfnamefont {J.-F.}\ \bibnamefont {Jia}},\ }\bibfield  {title} {\bibinfo {title} {Superconductivity above 100 k in single-layer fese films on doped srtio3},\ }\href@noop {} {\bibfield  {journal} {\bibinfo  {journal} {Nature Mater.}\ }\textbf {\bibinfo {volume} {14}},\ \bibinfo {pages} {285} (\bibinfo {year} {2015})}\BibitemShut {NoStop}%
\bibitem [{\citenamefont {Song}\ \emph {et~al.}(2019)\citenamefont {Song}, \citenamefont {Yu}, \citenamefont {Lou}, \citenamefont {Xie}, \citenamefont {Xu}, \citenamefont {Wen}, \citenamefont {Yao}, \citenamefont {Zhang}, \citenamefont {Zhu}, \citenamefont {Guo}, \citenamefont {Peng},\ and\ \citenamefont {Feng}}]{FeSe-STO2}%
  \BibitemOpen
  \bibfield  {author} {\bibinfo {author} {\bibfnamefont {Q.}~\bibnamefont {Song}}, \bibinfo {author} {\bibfnamefont {T.~L.}\ \bibnamefont {Yu}}, \bibinfo {author} {\bibfnamefont {X.}~\bibnamefont {Lou}}, \bibinfo {author} {\bibfnamefont {B.~P.}\ \bibnamefont {Xie}}, \bibinfo {author} {\bibfnamefont {H.~C.}\ \bibnamefont {Xu}}, \bibinfo {author} {\bibfnamefont {C.~H.~P.}\ \bibnamefont {Wen}}, \bibinfo {author} {\bibfnamefont {Q.}~\bibnamefont {Yao}}, \bibinfo {author} {\bibfnamefont {S.~Y.}\ \bibnamefont {Zhang}}, \bibinfo {author} {\bibfnamefont {X.~T.}\ \bibnamefont {Zhu}}, \bibinfo {author} {\bibfnamefont {J.~D.}\ \bibnamefont {Guo}}, \bibinfo {author} {\bibfnamefont {R.}~\bibnamefont {Peng}},\ and\ \bibinfo {author} {\bibfnamefont {D.~L.}\ \bibnamefont {Feng}},\ }\bibfield  {title} {\bibinfo {title} {Evidence of cooperative effect on the enhanced superconducting transition temperature at the fese/srtio3 interface},\ }\href {https://doi.org/10.1038/s41467-019-08560-z} {\bibfield  {journal} {\bibinfo
  {journal} {Nature Communications}\ }\textbf {\bibinfo {volume} {10}},\ \bibinfo {pages} {758} (\bibinfo {year} {2019})}\BibitemShut {NoStop}%
\bibitem [{\citenamefont {Lee}\ \emph {et~al.}(2014{\natexlab{a}})\citenamefont {Lee}, \citenamefont {Schmitt}, \citenamefont {Moore}, \citenamefont {Johnston}, \citenamefont {Cui}, \citenamefont {Li}, \citenamefont {Yi}, \citenamefont {Liu}, \citenamefont {Hashimoto}, \citenamefont {Zhang}, \citenamefont {Lu}, \citenamefont {Devereaux}, \citenamefont {Lee},\ and\ \citenamefont {Shen}}]{FeSe-STO3}%
  \BibitemOpen
  \bibfield  {author} {\bibinfo {author} {\bibfnamefont {J.~J.}\ \bibnamefont {Lee}}, \bibinfo {author} {\bibfnamefont {F.~T.}\ \bibnamefont {Schmitt}}, \bibinfo {author} {\bibfnamefont {R.~G.}\ \bibnamefont {Moore}}, \bibinfo {author} {\bibfnamefont {S.}~\bibnamefont {Johnston}}, \bibinfo {author} {\bibfnamefont {Y.~T.}\ \bibnamefont {Cui}}, \bibinfo {author} {\bibfnamefont {W.}~\bibnamefont {Li}}, \bibinfo {author} {\bibfnamefont {M.}~\bibnamefont {Yi}}, \bibinfo {author} {\bibfnamefont {Z.~K.}\ \bibnamefont {Liu}}, \bibinfo {author} {\bibfnamefont {M.}~\bibnamefont {Hashimoto}}, \bibinfo {author} {\bibfnamefont {Y.}~\bibnamefont {Zhang}}, \bibinfo {author} {\bibfnamefont {D.~H.}\ \bibnamefont {Lu}}, \bibinfo {author} {\bibfnamefont {T.~P.}\ \bibnamefont {Devereaux}}, \bibinfo {author} {\bibfnamefont {D.~H.}\ \bibnamefont {Lee}},\ and\ \bibinfo {author} {\bibfnamefont {Z.~X.}\ \bibnamefont {Shen}},\ }\bibfield  {title} {\bibinfo {title} {Interfacial mode coupling as the origin of the enhancement of tc in
  fese films on srtio3},\ }\href {https://doi.org/10.1038/nature13894} {\bibfield  {journal} {\bibinfo  {journal} {Nature}\ }\textbf {\bibinfo {volume} {515}},\ \bibinfo {pages} {245} (\bibinfo {year} {2014}{\natexlab{a}})}\BibitemShut {NoStop}%
\bibitem [{\citenamefont {Yang}\ \emph {et~al.}(2024)\citenamefont {Yang}, \citenamefont {Zhou}, \citenamefont {Miao}, \citenamefont {Rusz}, \citenamefont {Yan}, \citenamefont {Guzman}, \citenamefont {Xu}, \citenamefont {Xu}, \citenamefont {Aoki}, \citenamefont {Zeiger}, \citenamefont {Zhu}, \citenamefont {Wang}, \citenamefont {Guo}, \citenamefont {Wu},\ and\ \citenamefont {Pan}}]{EPC-nature}%
  \BibitemOpen
  \bibfield  {author} {\bibinfo {author} {\bibfnamefont {H.}~\bibnamefont {Yang}}, \bibinfo {author} {\bibfnamefont {Y.}~\bibnamefont {Zhou}}, \bibinfo {author} {\bibfnamefont {G.}~\bibnamefont {Miao}}, \bibinfo {author} {\bibfnamefont {J.}~\bibnamefont {Rusz}}, \bibinfo {author} {\bibfnamefont {X.}~\bibnamefont {Yan}}, \bibinfo {author} {\bibfnamefont {F.}~\bibnamefont {Guzman}}, \bibinfo {author} {\bibfnamefont {X.}~\bibnamefont {Xu}}, \bibinfo {author} {\bibfnamefont {X.}~\bibnamefont {Xu}}, \bibinfo {author} {\bibfnamefont {T.}~\bibnamefont {Aoki}}, \bibinfo {author} {\bibfnamefont {P.}~\bibnamefont {Zeiger}}, \bibinfo {author} {\bibfnamefont {X.}~\bibnamefont {Zhu}}, \bibinfo {author} {\bibfnamefont {W.}~\bibnamefont {Wang}}, \bibinfo {author} {\bibfnamefont {J.}~\bibnamefont {Guo}}, \bibinfo {author} {\bibfnamefont {R.}~\bibnamefont {Wu}},\ and\ \bibinfo {author} {\bibfnamefont {X.}~\bibnamefont {Pan}},\ }\bibfield  {title} {\bibinfo {title} {Phonon modes and electron--phonon coupling at the fese/srtio3
  interface},\ }\href {https://doi.org/10.1038/s41586-024-08118-0} {\bibfield  {journal} {\bibinfo  {journal} {Nature}\ }\textbf {\bibinfo {volume} {635}},\ \bibinfo {pages} {332} (\bibinfo {year} {2024})}\BibitemShut {NoStop}%
\bibitem [{\citenamefont {Coh}\ \emph {et~al.}(2015)\citenamefont {Coh}, \citenamefont {Cohen},\ and\ \citenamefont {Louie}}]{Coh2015}%
  \BibitemOpen
  \bibfield  {author} {\bibinfo {author} {\bibfnamefont {S.}~\bibnamefont {Coh}}, \bibinfo {author} {\bibfnamefont {M.~L.}\ \bibnamefont {Cohen}},\ and\ \bibinfo {author} {\bibfnamefont {S.~G.}\ \bibnamefont {Louie}},\ }\bibfield  {title} {\bibinfo {title} {Large electron–phonon interactions from fese phonons in a monolayer},\ }\href {https://doi.org/10.1088/1367-2630/17/7/073027} {\bibfield  {journal} {\bibinfo  {journal} {New Journal of Physics}\ }\textbf {\bibinfo {volume} {17}},\ \bibinfo {pages} {073027} (\bibinfo {year} {2015})}\BibitemShut {NoStop}%
\bibitem [{\citenamefont {Kim}\ \emph {et~al.}(2012)\citenamefont {Kim}, \citenamefont {Pashkin}, \citenamefont {Sch{\"a}fer}, \citenamefont {Beyer}, \citenamefont {Porer}, \citenamefont {Wolf}, \citenamefont {Bernhard}, \citenamefont {Demsar}, \citenamefont {Huber},\ and\ \citenamefont {Leitenstorfer}}]{EPC-Ba122}%
  \BibitemOpen
  \bibfield  {author} {\bibinfo {author} {\bibfnamefont {K.~W.}\ \bibnamefont {Kim}}, \bibinfo {author} {\bibfnamefont {A.}~\bibnamefont {Pashkin}}, \bibinfo {author} {\bibfnamefont {H.}~\bibnamefont {Sch{\"a}fer}}, \bibinfo {author} {\bibfnamefont {M.}~\bibnamefont {Beyer}}, \bibinfo {author} {\bibfnamefont {M.}~\bibnamefont {Porer}}, \bibinfo {author} {\bibfnamefont {T.}~\bibnamefont {Wolf}}, \bibinfo {author} {\bibfnamefont {C.}~\bibnamefont {Bernhard}}, \bibinfo {author} {\bibfnamefont {J.}~\bibnamefont {Demsar}}, \bibinfo {author} {\bibfnamefont {R.}~\bibnamefont {Huber}},\ and\ \bibinfo {author} {\bibfnamefont {A.}~\bibnamefont {Leitenstorfer}},\ }\bibfield  {title} {\bibinfo {title} {Ultrafast transient generation of spin-density-wave order in the normal state of bafe2as2 driven by coherent lattice vibrations},\ }\href {https://doi.org/10.1038/nmat3294} {\bibfield  {journal} {\bibinfo  {journal} {Nature Materials}\ }\textbf {\bibinfo {volume} {11}},\ \bibinfo {pages} {497} (\bibinfo {year}
  {2012})}\BibitemShut {NoStop}%
\bibitem [{\citenamefont {Khanal}\ and\ \citenamefont {Haule}(2020)}]{phonon-KH}%
  \BibitemOpen
  \bibfield  {author} {\bibinfo {author} {\bibfnamefont {G.}~\bibnamefont {Khanal}}\ and\ \bibinfo {author} {\bibfnamefont {K.}~\bibnamefont {Haule}},\ }\bibfield  {title} {\bibinfo {title} {Correlation driven phonon anomalies in bulk fese},\ }\href {https://doi.org/10.1103/PhysRevB.102.241108} {\bibfield  {journal} {\bibinfo  {journal} {Phys. Rev. B}\ }\textbf {\bibinfo {volume} {102}},\ \bibinfo {pages} {241108} (\bibinfo {year} {2020})}\BibitemShut {NoStop}%
\bibitem [{\citenamefont {Okabe}\ \emph {et~al.}(2010)\citenamefont {Okabe}, \citenamefont {Takeshita}, \citenamefont {Horigane}, \citenamefont {Muranaka},\ and\ \citenamefont {Akimitsu}}]{angle-tc-2}%
  \BibitemOpen
  \bibfield  {author} {\bibinfo {author} {\bibfnamefont {H.}~\bibnamefont {Okabe}}, \bibinfo {author} {\bibfnamefont {N.}~\bibnamefont {Takeshita}}, \bibinfo {author} {\bibfnamefont {K.}~\bibnamefont {Horigane}}, \bibinfo {author} {\bibfnamefont {T.}~\bibnamefont {Muranaka}},\ and\ \bibinfo {author} {\bibfnamefont {J.}~\bibnamefont {Akimitsu}},\ }\bibfield  {title} {\bibinfo {title} {Pressure-induced high-${T}_{c}$ superconducting phase in fese: Correlation between anion height and ${T}_{c}$},\ }\href {https://doi.org/10.1103/PhysRevB.81.205119} {\bibfield  {journal} {\bibinfo  {journal} {Phys. Rev. B}\ }\textbf {\bibinfo {volume} {81}},\ \bibinfo {pages} {205119} (\bibinfo {year} {2010})}\BibitemShut {NoStop}%
\bibitem [{\citenamefont {Hosono}\ \emph {et~al.}(2018)\citenamefont {Hosono}, \citenamefont {Yamamoto}, \citenamefont {Hiramatsu},\ and\ \citenamefont {Ma}}]{angle-tc}%
  \BibitemOpen
  \bibfield  {author} {\bibinfo {author} {\bibfnamefont {H.}~\bibnamefont {Hosono}}, \bibinfo {author} {\bibfnamefont {A.}~\bibnamefont {Yamamoto}}, \bibinfo {author} {\bibfnamefont {H.}~\bibnamefont {Hiramatsu}},\ and\ \bibinfo {author} {\bibfnamefont {Y.}~\bibnamefont {Ma}},\ }\bibfield  {title} {\bibinfo {title} {Recent advances in iron-based superconductors toward applications},\ }\href {https://doi.org/https://doi.org/10.1016/j.mattod.2017.09.006} {\bibfield  {journal} {\bibinfo  {journal} {Materials Today}\ }\textbf {\bibinfo {volume} {21}},\ \bibinfo {pages} {278} (\bibinfo {year} {2018})}\BibitemShut {NoStop}%
\bibitem [{\citenamefont {Keimer}\ \emph {et~al.}(2015)\citenamefont {Keimer}, \citenamefont {Kivelson}, \citenamefont {Norman}, \citenamefont {Uchida},\ and\ \citenamefont {Zaanen}}]{Keimer2015}%
  \BibitemOpen
  \bibfield  {author} {\bibinfo {author} {\bibfnamefont {B.}~\bibnamefont {Keimer}}, \bibinfo {author} {\bibfnamefont {S.~A.}\ \bibnamefont {Kivelson}}, \bibinfo {author} {\bibfnamefont {M.~R.}\ \bibnamefont {Norman}}, \bibinfo {author} {\bibfnamefont {S.}~\bibnamefont {Uchida}},\ and\ \bibinfo {author} {\bibfnamefont {J.}~\bibnamefont {Zaanen}},\ }\bibfield  {title} {\bibinfo {title} {From quantum matter to high-temperature superconductivity in copper oxides},\ }\href@noop {} {\bibfield  {journal} {\bibinfo  {journal} {Nature}\ }\textbf {\bibinfo {volume} {518}},\ \bibinfo {pages} {179} (\bibinfo {year} {2015})}\BibitemShut {NoStop}%
\bibitem [{\citenamefont {Fernandes}\ \emph {et~al.}(2014)\citenamefont {Fernandes}, \citenamefont {Chubukov},\ and\ \citenamefont {Schmalian}}]{nematic1}%
  \BibitemOpen
  \bibfield  {author} {\bibinfo {author} {\bibfnamefont {R.~M.}\ \bibnamefont {Fernandes}}, \bibinfo {author} {\bibfnamefont {A.~V.}\ \bibnamefont {Chubukov}},\ and\ \bibinfo {author} {\bibfnamefont {J.}~\bibnamefont {Schmalian}},\ }\bibfield  {title} {\bibinfo {title} {What drives nematic order in iron-based superconductors?},\ }\href {https://doi.org/10.1038/nphys2877} {\bibfield  {journal} {\bibinfo  {journal} {Nature Physics}\ }\textbf {\bibinfo {volume} {10}},\ \bibinfo {pages} {97} (\bibinfo {year} {2014})}\BibitemShut {NoStop}%
\bibitem [{\citenamefont {Shimojima}\ \emph {et~al.}(2021)\citenamefont {Shimojima}, \citenamefont {Motoyui}, \citenamefont {Taniuchi}, \citenamefont {Bareille}, \citenamefont {Onari}, \citenamefont {Kontani}, \citenamefont {Nakajima}, \citenamefont {Kasahara}, \citenamefont {Shibauchi}, \citenamefont {Matsuda},\ and\ \citenamefont {Shin}}]{nematic2}%
  \BibitemOpen
  \bibfield  {author} {\bibinfo {author} {\bibfnamefont {T.}~\bibnamefont {Shimojima}}, \bibinfo {author} {\bibfnamefont {Y.}~\bibnamefont {Motoyui}}, \bibinfo {author} {\bibfnamefont {T.}~\bibnamefont {Taniuchi}}, \bibinfo {author} {\bibfnamefont {C.}~\bibnamefont {Bareille}}, \bibinfo {author} {\bibfnamefont {S.}~\bibnamefont {Onari}}, \bibinfo {author} {\bibfnamefont {H.}~\bibnamefont {Kontani}}, \bibinfo {author} {\bibfnamefont {M.}~\bibnamefont {Nakajima}}, \bibinfo {author} {\bibfnamefont {S.}~\bibnamefont {Kasahara}}, \bibinfo {author} {\bibfnamefont {T.}~\bibnamefont {Shibauchi}}, \bibinfo {author} {\bibfnamefont {Y.}~\bibnamefont {Matsuda}},\ and\ \bibinfo {author} {\bibfnamefont {S.}~\bibnamefont {Shin}},\ }\bibfield  {title} {\bibinfo {title} {Discovery of mesoscopic nematicity wave in iron-based superconductors},\ }\href {https://doi.org/10.1126/science.abd6701} {\bibfield  {journal} {\bibinfo  {journal} {Science}\ }\textbf {\bibinfo {volume} {373}},\ \bibinfo {pages} {1122} (\bibinfo {year}
  {2021})},\ \Eprint {https://arxiv.org/abs/https://www.science.org/doi/pdf/10.1126/science.abd6701} {https://www.science.org/doi/pdf/10.1126/science.abd6701} \BibitemShut {NoStop}%
\bibitem [{\citenamefont {Nag}\ \emph {et~al.}(2025)\citenamefont {Nag}, \citenamefont {Scott}, \citenamefont {de~Carvalho}, \citenamefont {Byland}, \citenamefont {Yang}, \citenamefont {Walker}, \citenamefont {Greenberg}, \citenamefont {Klavins}, \citenamefont {Miranda}, \citenamefont {Gozar}, \citenamefont {Taufour}, \citenamefont {Fernandes},\ and\ \citenamefont {da~Silva~Neto}}]{nematic3}%
  \BibitemOpen
  \bibfield  {author} {\bibinfo {author} {\bibfnamefont {P.~K.}\ \bibnamefont {Nag}}, \bibinfo {author} {\bibfnamefont {K.}~\bibnamefont {Scott}}, \bibinfo {author} {\bibfnamefont {V.~S.}\ \bibnamefont {de~Carvalho}}, \bibinfo {author} {\bibfnamefont {J.~K.}\ \bibnamefont {Byland}}, \bibinfo {author} {\bibfnamefont {X.}~\bibnamefont {Yang}}, \bibinfo {author} {\bibfnamefont {M.}~\bibnamefont {Walker}}, \bibinfo {author} {\bibfnamefont {A.~G.}\ \bibnamefont {Greenberg}}, \bibinfo {author} {\bibfnamefont {P.}~\bibnamefont {Klavins}}, \bibinfo {author} {\bibfnamefont {E.}~\bibnamefont {Miranda}}, \bibinfo {author} {\bibfnamefont {A.}~\bibnamefont {Gozar}}, \bibinfo {author} {\bibfnamefont {V.}~\bibnamefont {Taufour}}, \bibinfo {author} {\bibfnamefont {R.~M.}\ \bibnamefont {Fernandes}},\ and\ \bibinfo {author} {\bibfnamefont {E.~H.}\ \bibnamefont {da~Silva~Neto}},\ }\bibfield  {title} {\bibinfo {title} {Highly anisotropic superconducting gap near the nematic quantum critical point of fese1-xsx},\ }\href
  {https://doi.org/10.1038/s41567-024-02683-x} {\bibfield  {journal} {\bibinfo  {journal} {Nature Physics}\ }\textbf {\bibinfo {volume} {21}},\ \bibinfo {pages} {89} (\bibinfo {year} {2025})}\BibitemShut {NoStop}%
\bibitem [{\citenamefont {Lou}\ \emph {et~al.}(2023)\citenamefont {Lou}, \citenamefont {Suvorov}, \citenamefont {Grafe}, \citenamefont {Kuibarov}, \citenamefont {Krivenkov}, \citenamefont {Rader}, \citenamefont {B{\"u}chner}, \citenamefont {Borisenko},\ and\ \citenamefont {Fedorov}}]{SLFeSe_nematic_suppression}%
  \BibitemOpen
  \bibfield  {author} {\bibinfo {author} {\bibfnamefont {R.}~\bibnamefont {Lou}}, \bibinfo {author} {\bibfnamefont {O.}~\bibnamefont {Suvorov}}, \bibinfo {author} {\bibfnamefont {H.-J.}\ \bibnamefont {Grafe}}, \bibinfo {author} {\bibfnamefont {A.}~\bibnamefont {Kuibarov}}, \bibinfo {author} {\bibfnamefont {M.}~\bibnamefont {Krivenkov}}, \bibinfo {author} {\bibfnamefont {O.}~\bibnamefont {Rader}}, \bibinfo {author} {\bibfnamefont {B.}~\bibnamefont {B{\"u}chner}}, \bibinfo {author} {\bibfnamefont {S.}~\bibnamefont {Borisenko}},\ and\ \bibinfo {author} {\bibfnamefont {A.}~\bibnamefont {Fedorov}},\ }\bibfield  {title} {\bibinfo {title} {Suppression of nematicity by tensile strain in multilayer fese/srtio$_3$ films},\ }\href {https://doi.org/10.1103/PhysRevResearch.5.043011} {\bibfield  {journal} {\bibinfo  {journal} {Physical Review Research}\ }\textbf {\bibinfo {volume} {5}},\ \bibinfo {pages} {043011} (\bibinfo {year} {2023})}\BibitemShut {NoStop}%
\bibitem [{\citenamefont {Yang}\ \emph {et~al.}(2019)\citenamefont {Yang}, \citenamefont {Yan}, \citenamefont {Ma}, \citenamefont {Li},\ and\ \citenamefont {Cen}}]{UV-Lian}%
  \BibitemOpen
  \bibfield  {author} {\bibinfo {author} {\bibfnamefont {M.}~\bibnamefont {Yang}}, \bibinfo {author} {\bibfnamefont {C.}~\bibnamefont {Yan}}, \bibinfo {author} {\bibfnamefont {Y.}~\bibnamefont {Ma}}, \bibinfo {author} {\bibfnamefont {L.}~\bibnamefont {Li}},\ and\ \bibinfo {author} {\bibfnamefont {C.}~\bibnamefont {Cen}},\ }\bibfield  {title} {\bibinfo {title} {Light induced non-volatile switching of superconductivity in single layer fese on srtio3 substrate},\ }\href {https://doi.org/10.1038/s41467-018-08024-w} {\bibfield  {journal} {\bibinfo  {journal} {Nature Communications}\ }\textbf {\bibinfo {volume} {10}},\ \bibinfo {pages} {85} (\bibinfo {year} {2019})}\BibitemShut {NoStop}%
\bibitem [{\citenamefont {Zou}\ \emph {et~al.}(2023)\citenamefont {Zou}, \citenamefont {Oli}, \citenamefont {Zhang}, \citenamefont {Benigno}, \citenamefont {Li},\ and\ \citenamefont {Li}}]{Lian-3}%
  \BibitemOpen
  \bibfield  {author} {\bibinfo {author} {\bibfnamefont {Q.}~\bibnamefont {Zou}}, \bibinfo {author} {\bibfnamefont {B.~D.}\ \bibnamefont {Oli}}, \bibinfo {author} {\bibfnamefont {H.}~\bibnamefont {Zhang}}, \bibinfo {author} {\bibfnamefont {J.}~\bibnamefont {Benigno}}, \bibinfo {author} {\bibfnamefont {X.}~\bibnamefont {Li}},\ and\ \bibinfo {author} {\bibfnamefont {L.}~\bibnamefont {Li}},\ }\bibfield  {title} {\bibinfo {title} {Deciphering alloy composition in superconducting single-layer fese1--xsx on srtio3(001) substrates by machine learning of stm/s data},\ }\href {https://doi.org/10.1021/acsami.2c23324} {\bibfield  {journal} {\bibinfo  {journal} {ACS Applied Materials \& Interfaces}\ }\textbf {\bibinfo {volume} {15}},\ \bibinfo {pages} {22644} (\bibinfo {year} {2023})}\BibitemShut {NoStop}%
\bibitem [{\citenamefont {Oli}\ \emph {et~al.}(2023)\citenamefont {Oli}, \citenamefont {Zou}, \citenamefont {Li},\ and\ \citenamefont {Li}}]{Lian-4}%
  \BibitemOpen
  \bibfield  {author} {\bibinfo {author} {\bibfnamefont {B.~D.}\ \bibnamefont {Oli}}, \bibinfo {author} {\bibfnamefont {Q.}~\bibnamefont {Zou}}, \bibinfo {author} {\bibfnamefont {X.}~\bibnamefont {Li}},\ and\ \bibinfo {author} {\bibfnamefont {L.}~\bibnamefont {Li}},\ }\bibfield  {title} {\bibinfo {title} {Atomic-scale electronic inhomogeneity in single-layer iron chalcogenide alloys revealed by machine learning of stm/s data},\ }\href {https://doi.org/10.1063/5.0156695} {\bibfield  {journal} {\bibinfo  {journal} {AIP Advances}\ }\textbf {\bibinfo {volume} {13}},\ \bibinfo {pages} {105224} (\bibinfo {year} {2023})}\BibitemShut {NoStop}%
\bibitem [{\citenamefont {Song}\ \emph {et~al.}(2011)\citenamefont {Song}, \citenamefont {Wang}, \citenamefont {Cheng}, \citenamefont {Jiang}, \citenamefont {Li}, \citenamefont {Zhang}, \citenamefont {Li}, \citenamefont {He}, \citenamefont {Wang}, \citenamefont {Jia}, \citenamefont {Hung}, \citenamefont {Wu}, \citenamefont {Ma}, \citenamefont {Chen},\ and\ \citenamefont {Xue}}]{doi:10.1126/science.1202226}%
  \BibitemOpen
  \bibfield  {author} {\bibinfo {author} {\bibfnamefont {C.-L.}\ \bibnamefont {Song}}, \bibinfo {author} {\bibfnamefont {Y.-L.}\ \bibnamefont {Wang}}, \bibinfo {author} {\bibfnamefont {P.}~\bibnamefont {Cheng}}, \bibinfo {author} {\bibfnamefont {Y.-P.}\ \bibnamefont {Jiang}}, \bibinfo {author} {\bibfnamefont {W.}~\bibnamefont {Li}}, \bibinfo {author} {\bibfnamefont {T.}~\bibnamefont {Zhang}}, \bibinfo {author} {\bibfnamefont {Z.}~\bibnamefont {Li}}, \bibinfo {author} {\bibfnamefont {K.}~\bibnamefont {He}}, \bibinfo {author} {\bibfnamefont {L.}~\bibnamefont {Wang}}, \bibinfo {author} {\bibfnamefont {J.-F.}\ \bibnamefont {Jia}}, \bibinfo {author} {\bibfnamefont {H.-H.}\ \bibnamefont {Hung}}, \bibinfo {author} {\bibfnamefont {C.}~\bibnamefont {Wu}}, \bibinfo {author} {\bibfnamefont {X.}~\bibnamefont {Ma}}, \bibinfo {author} {\bibfnamefont {X.}~\bibnamefont {Chen}},\ and\ \bibinfo {author} {\bibfnamefont {Q.-K.}\ \bibnamefont {Xue}},\ }\bibfield  {title} {\bibinfo {title} {Direct observation of nodes and twofold
  symmetry in fese superconductor},\ }\href {https://doi.org/10.1126/science.1202226} {\bibfield  {journal} {\bibinfo  {journal} {Science}\ }\textbf {\bibinfo {volume} {332}},\ \bibinfo {pages} {1410} (\bibinfo {year} {2011})},\ \Eprint {https://arxiv.org/abs/https://www.science.org/doi/pdf/10.1126/science.1202226} {https://www.science.org/doi/pdf/10.1126/science.1202226} \BibitemShut {NoStop}%
\bibitem [{\citenamefont {Haule}\ and\ \citenamefont {Pascut}(2016)}]{Forces}%
  \BibitemOpen
  \bibfield  {author} {\bibinfo {author} {\bibfnamefont {K.}~\bibnamefont {Haule}}\ and\ \bibinfo {author} {\bibfnamefont {G.~L.}\ \bibnamefont {Pascut}},\ }\bibfield  {title} {\bibinfo {title} {Forces for structural optimizations in correlated materials within a dft+embedded dmft functional approach},\ }\href {https://doi.org/10.1103/PhysRevB.94.195146} {\bibfield  {journal} {\bibinfo  {journal} {Phys. Rev. B}\ }\textbf {\bibinfo {volume} {94}},\ \bibinfo {pages} {195146} (\bibinfo {year} {2016})}\BibitemShut {NoStop}%
\bibitem [{\citenamefont {Peng}\ \emph {et~al.}(2020)\citenamefont {Peng}, \citenamefont {Zou}, \citenamefont {Han}, \citenamefont {Albright}, \citenamefont {Hong}, \citenamefont {Lau}, \citenamefont {Xu}, \citenamefont {Zhu}, \citenamefont {Walker},\ and\ \citenamefont {Ahn}}]{charles-sciadv}%
  \BibitemOpen
  \bibfield  {author} {\bibinfo {author} {\bibfnamefont {R.}~\bibnamefont {Peng}}, \bibinfo {author} {\bibfnamefont {K.}~\bibnamefont {Zou}}, \bibinfo {author} {\bibfnamefont {M.~G.}\ \bibnamefont {Han}}, \bibinfo {author} {\bibfnamefont {S.~D.}\ \bibnamefont {Albright}}, \bibinfo {author} {\bibfnamefont {H.}~\bibnamefont {Hong}}, \bibinfo {author} {\bibfnamefont {C.}~\bibnamefont {Lau}}, \bibinfo {author} {\bibfnamefont {H.~C.}\ \bibnamefont {Xu}}, \bibinfo {author} {\bibfnamefont {Y.}~\bibnamefont {Zhu}}, \bibinfo {author} {\bibfnamefont {F.~J.}\ \bibnamefont {Walker}},\ and\ \bibinfo {author} {\bibfnamefont {C.~H.}\ \bibnamefont {Ahn}},\ }\bibfield  {title} {\bibinfo {title} {Picoscale structural insight into superconductivity of monolayer fese/srtio<sub>3</sub>},\ }\href {https://doi.org/10.1126/sciadv.aay4517} {\bibfield  {journal} {\bibinfo  {journal} {Science Advances}\ }\textbf {\bibinfo {volume} {6}},\ \bibinfo {pages} {eaay4517} (\bibinfo {year} {2020})},\ \Eprint
  {https://arxiv.org/abs/https://www.science.org/doi/pdf/10.1126/sciadv.aay4517} {https://www.science.org/doi/pdf/10.1126/sciadv.aay4517} \BibitemShut {NoStop}%
\bibitem [{\citenamefont {Zou}\ \emph {et~al.}(2025)\citenamefont {Zou}, \citenamefont {Kim}, \citenamefont {Kang}, \citenamefont {Oli}, \citenamefont {Ge}, \citenamefont {Weinert}, \citenamefont {Mandal}, \citenamefont {Eom}, \citenamefont {Choi},\ and\ \citenamefont {Li}}]{substrate-Mandal}%
  \BibitemOpen
  \bibfield  {author} {\bibinfo {author} {\bibfnamefont {Q.}~\bibnamefont {Zou}}, \bibinfo {author} {\bibfnamefont {G.-Y.}\ \bibnamefont {Kim}}, \bibinfo {author} {\bibfnamefont {J.-H.}\ \bibnamefont {Kang}}, \bibinfo {author} {\bibfnamefont {B.~D.}\ \bibnamefont {Oli}}, \bibinfo {author} {\bibfnamefont {Z.}~\bibnamefont {Ge}}, \bibinfo {author} {\bibfnamefont {M.}~\bibnamefont {Weinert}}, \bibinfo {author} {\bibfnamefont {S.}~\bibnamefont {Mandal}}, \bibinfo {author} {\bibfnamefont {C.-B.}\ \bibnamefont {Eom}}, \bibinfo {author} {\bibfnamefont {S.-Y.}\ \bibnamefont {Choi}},\ and\ \bibinfo {author} {\bibfnamefont {L.}~\bibnamefont {Li}},\ }\bibfield  {title} {\bibinfo {title} {Unraveling enhanced superconductivity in single-layer fese through substrate surface terminations},\ }\href {https://arxiv.org/abs/2502.16784} {\  (\bibinfo {year} {2025})},\ \Eprint {https://arxiv.org/abs/2502.16784} {arXiv:2502.16784 [cond-mat.supr-con]} \BibitemShut {NoStop}%
\bibitem [{\citenamefont {Haule}\ and\ \citenamefont {Kotliar}(2009)}]{Haule-njp09}%
  \BibitemOpen
  \bibfield  {author} {\bibinfo {author} {\bibfnamefont {K.}~\bibnamefont {Haule}}\ and\ \bibinfo {author} {\bibfnamefont {G.}~\bibnamefont {Kotliar}},\ }\bibfield  {title} {\bibinfo {title} {Coherence-incoherence crossover in the normal state of iron oxypnictides and importance of hund's rule coupling},\ }\href@noop {} {\bibfield  {journal} {\bibinfo  {journal} {New Journal of Physics}\ }\textbf {\bibinfo {volume} {11}},\ \bibinfo {pages} {025021} (\bibinfo {year} {2009})}\BibitemShut {NoStop}%
\bibitem [{\citenamefont {Yin}\ \emph {et~al.}(2011)\citenamefont {Yin}, \citenamefont {Haule},\ and\ \citenamefont {Kotliar}}]{yin2011kinetic}%
  \BibitemOpen
  \bibfield  {author} {\bibinfo {author} {\bibfnamefont {Z.}~\bibnamefont {Yin}}, \bibinfo {author} {\bibfnamefont {K.}~\bibnamefont {Haule}},\ and\ \bibinfo {author} {\bibfnamefont {G.}~\bibnamefont {Kotliar}},\ }\bibfield  {title} {\bibinfo {title} {Kinetic frustration and the nature of the magnetic and paramagnetic states in iron pnictides and iron chalcogenides},\ }\href@noop {} {\bibfield  {journal} {\bibinfo  {journal} {Nature materials}\ }\textbf {\bibinfo {volume} {10}},\ \bibinfo {pages} {932} (\bibinfo {year} {2011})}\BibitemShut {NoStop}%
\bibitem [{\citenamefont {Mandal}\ \emph {et~al.}(2014{\natexlab{b}})\citenamefont {Mandal}, \citenamefont {Cohen},\ and\ \citenamefont {Haule}}]{Mandal:2014}%
  \BibitemOpen
  \bibfield  {author} {\bibinfo {author} {\bibfnamefont {S.}~\bibnamefont {Mandal}}, \bibinfo {author} {\bibfnamefont {R.~E.}\ \bibnamefont {Cohen}},\ and\ \bibinfo {author} {\bibfnamefont {K.}~\bibnamefont {Haule}},\ }\bibfield  {title} {\bibinfo {title} {Pressure suppression of electron correlation in the collapsed tetragonal phase of ${\mathrm{cafe}}_{2}{\mathrm{as}}_{2}$: A dft-dmft investigation},\ }\href {https://doi.org/10.1103/PhysRevB.90.060501} {\bibfield  {journal} {\bibinfo  {journal} {Phys. Rev. B}\ }\textbf {\bibinfo {volume} {90}},\ \bibinfo {pages} {060501} (\bibinfo {year} {2014}{\natexlab{b}})}\BibitemShut {NoStop}%
\bibitem [{\citenamefont {Backes}\ \emph {et~al.}(2015)\citenamefont {Backes}, \citenamefont {Jeschke},\ and\ \citenamefont {Valentí}}]{Roser2015}%
  \BibitemOpen
  \bibfield  {author} {\bibinfo {author} {\bibfnamefont {S.}~\bibnamefont {Backes}}, \bibinfo {author} {\bibfnamefont {H.~O.}\ \bibnamefont {Jeschke}},\ and\ \bibinfo {author} {\bibfnamefont {R.}~\bibnamefont {Valentí}},\ }\bibfield  {title} {\bibinfo {title} {Microscopic nature of correlations in multi-orbital afe$_2$as$_2$ (a=k, rb, cs): Hund’s coupling versus coulomb repulsion},\ }\href {https://doi.org/10.1103/PhysRevB.92.195128} {\bibfield  {journal} {\bibinfo  {journal} {Physical Review B}\ }\textbf {\bibinfo {volume} {92}},\ \bibinfo {pages} {195128} (\bibinfo {year} {2015})}\BibitemShut {NoStop}%
\bibitem [{\citenamefont {Georges}\ \emph {et~al.}(2013)\citenamefont {Georges}, \citenamefont {Medici},\ and\ \citenamefont {Mravlje}}]{luca-1}%
  \BibitemOpen
  \bibfield  {author} {\bibinfo {author} {\bibfnamefont {A.}~\bibnamefont {Georges}}, \bibinfo {author} {\bibfnamefont {L.~d.}\ \bibnamefont {Medici}},\ and\ \bibinfo {author} {\bibfnamefont {J.}~\bibnamefont {Mravlje}},\ }\bibfield  {title} {\bibinfo {title} {Strong correlations from hund's coupling},\ }\href {https://doi.org/https://doi.org/10.1146/annurev-conmatphys-020911-125045} {\bibfield  {journal} {\bibinfo  {journal} {Annual Review of Condensed Matter Physics}\ }\textbf {\bibinfo {volume} {4}},\ \bibinfo {pages} {137} (\bibinfo {year} {2013})}\BibitemShut {NoStop}%
\bibitem [{\citenamefont {Mandal}\ \emph {et~al.}(2017)\citenamefont {Mandal}, \citenamefont {Zhang}, \citenamefont {Ismail-Beigi},\ and\ \citenamefont {Haule}}]{mandal-PRL}%
  \BibitemOpen
  \bibfield  {author} {\bibinfo {author} {\bibfnamefont {S.}~\bibnamefont {Mandal}}, \bibinfo {author} {\bibfnamefont {P.}~\bibnamefont {Zhang}}, \bibinfo {author} {\bibfnamefont {S.}~\bibnamefont {Ismail-Beigi}},\ and\ \bibinfo {author} {\bibfnamefont {K.}~\bibnamefont {Haule}},\ }\bibfield  {title} {\bibinfo {title} {How correlated is the $\mathrm{FeSe}/{\mathrm{srtio}}_{3}$ system?},\ }\href {https://doi.org/10.1103/PhysRevLett.119.067004} {\bibfield  {journal} {\bibinfo  {journal} {Phys. Rev. Lett.}\ }\textbf {\bibinfo {volume} {119}},\ \bibinfo {pages} {067004} (\bibinfo {year} {2017})}\BibitemShut {NoStop}%
\bibitem [{\citenamefont {{Lin}}\ \emph {et~al.}(2025)\citenamefont {{Lin}}, \citenamefont {{Jacobs}}, \citenamefont {{Yan}}, \citenamefont {{Nolan}}, \citenamefont {{Berruto}}, \citenamefont {{Singleton}}, \citenamefont {{Nguyen}}, \citenamefont {{Bai}}, \citenamefont {{Gao}}, \citenamefont {{Wu}}, \citenamefont {{Liu}}, \citenamefont {{Yan}}, \citenamefont {{Choi}}, \citenamefont {{Liu}}, \citenamefont {{Guisinger}}, \citenamefont {{Huang}}, \citenamefont {{Mandal}},\ and\ \citenamefont {{Yang}}}]{Chris-arxiv}%
  \BibitemOpen
  \bibfield  {author} {\bibinfo {author} {\bibfnamefont {H.}~\bibnamefont {{Lin}}}, \bibinfo {author} {\bibfnamefont {C.~L.}\ \bibnamefont {{Jacobs}}}, \bibinfo {author} {\bibfnamefont {C.}~\bibnamefont {{Yan}}}, \bibinfo {author} {\bibfnamefont {G.~M.}\ \bibnamefont {{Nolan}}}, \bibinfo {author} {\bibfnamefont {G.}~\bibnamefont {{Berruto}}}, \bibinfo {author} {\bibfnamefont {P.}~\bibnamefont {{Singleton}}}, \bibinfo {author} {\bibfnamefont {K.~D.}\ \bibnamefont {{Nguyen}}}, \bibinfo {author} {\bibfnamefont {Y.}~\bibnamefont {{Bai}}}, \bibinfo {author} {\bibfnamefont {Q.}~\bibnamefont {{Gao}}}, \bibinfo {author} {\bibfnamefont {X.}~\bibnamefont {{Wu}}}, \bibinfo {author} {\bibfnamefont {C.-X.}\ \bibnamefont {{Liu}}}, \bibinfo {author} {\bibfnamefont {G.}~\bibnamefont {{Yan}}}, \bibinfo {author} {\bibfnamefont {S.}~\bibnamefont {{Choi}}}, \bibinfo {author} {\bibfnamefont {C.}~\bibnamefont {{Liu}}}, \bibinfo {author} {\bibfnamefont {N.~P.}\ \bibnamefont {{Guisinger}}}, \bibinfo {author} {\bibfnamefont {P.~Y.}\
  \bibnamefont {{Huang}}}, \bibinfo {author} {\bibfnamefont {S.}~\bibnamefont {{Mandal}}},\ and\ \bibinfo {author} {\bibfnamefont {S.}~\bibnamefont {{Yang}}},\ }\bibfield  {title} {\bibinfo {title} {{A Topological Superconductor Tuned by Electronic Correlations}},\ }\href {https://doi.org/10.48550/arXiv.2503.22888} {\bibfield  {journal} {\bibinfo  {journal} {arXiv e-prints}\ ,\ \bibinfo {eid} {arXiv:2503.22888}} (\bibinfo {year} {2025})},\ \Eprint {https://arxiv.org/abs/2503.22888} {arXiv:2503.22888 [cond-mat.supr-con]} \BibitemShut {NoStop}%
\bibitem [{\citenamefont {Gnezdilov}\ \emph {et~al.}(2013)\citenamefont {Gnezdilov}, \citenamefont {Pashkevich}, \citenamefont {Lemmens}, \citenamefont {Wulferding}, \citenamefont {Shevtsova}, \citenamefont {Gusev}, \citenamefont {Chareev},\ and\ \citenamefont {Vasiliev}}]{PhysRevB.87.144508}%
  \BibitemOpen
  \bibfield  {author} {\bibinfo {author} {\bibfnamefont {V.}~\bibnamefont {Gnezdilov}}, \bibinfo {author} {\bibfnamefont {Y.~G.}\ \bibnamefont {Pashkevich}}, \bibinfo {author} {\bibfnamefont {P.}~\bibnamefont {Lemmens}}, \bibinfo {author} {\bibfnamefont {D.}~\bibnamefont {Wulferding}}, \bibinfo {author} {\bibfnamefont {T.}~\bibnamefont {Shevtsova}}, \bibinfo {author} {\bibfnamefont {A.}~\bibnamefont {Gusev}}, \bibinfo {author} {\bibfnamefont {D.}~\bibnamefont {Chareev}},\ and\ \bibinfo {author} {\bibfnamefont {A.}~\bibnamefont {Vasiliev}},\ }\bibfield  {title} {\bibinfo {title} {Interplay between lattice and spin states degree of freedom in the fese superconductor: Dynamic spin state instabilities},\ }\href {https://doi.org/10.1103/PhysRevB.87.144508} {\bibfield  {journal} {\bibinfo  {journal} {Phys. Rev. B}\ }\textbf {\bibinfo {volume} {87}},\ \bibinfo {pages} {144508} (\bibinfo {year} {2013})}\BibitemShut {NoStop}%
\bibitem [{\citenamefont {Lazarevi\ifmmode~\acute{c}\else \'{c}\fi{}}\ \emph {et~al.}(2022)\citenamefont {Lazarevi\ifmmode~\acute{c}\else \'{c}\fi{}}, \citenamefont {Baum}, \citenamefont {Milosavljevi\ifmmode~\acute{c}\else \'{c}\fi{}}, \citenamefont {Peis}, \citenamefont {Stumberger}, \citenamefont {Bekaert}, \citenamefont {\ifmmode \check{S}\else \v{S}\fi{}olaji\ifmmode~\acute{c}\else \'{c}\fi{}}, \citenamefont {Pe\ifmmode \check{s}\else \v{s}\fi{}i\ifmmode~\acute{c}\else \'{c}\fi{}}, \citenamefont {Wang}, \citenamefont {\ifmmode \check{S}\else \v{S}\fi{}\ifmmode \acute{c}\else \'{c}\fi{}epanovi\ifmmode~\acute{c}\else \'{c}\fi{}}, \citenamefont {Abeykoon}, \citenamefont {Milo\ifmmode \check{s}\else \v{s}\fi{}evi\ifmmode~\acute{c}\else \'{c}\fi{}}, \citenamefont {Petrovic}, \citenamefont {Popovi\ifmmode~\acute{c}\else \'{c}\fi{}},\ and\ \citenamefont {Hackl}}]{FeSeS}%
  \BibitemOpen
  \bibfield  {author} {\bibinfo {author} {\bibfnamefont {N.}~\bibnamefont {Lazarevi\ifmmode~\acute{c}\else \'{c}\fi{}}}, \bibinfo {author} {\bibfnamefont {A.}~\bibnamefont {Baum}}, \bibinfo {author} {\bibfnamefont {A.}~\bibnamefont {Milosavljevi\ifmmode~\acute{c}\else \'{c}\fi{}}}, \bibinfo {author} {\bibfnamefont {L.}~\bibnamefont {Peis}}, \bibinfo {author} {\bibfnamefont {R.}~\bibnamefont {Stumberger}}, \bibinfo {author} {\bibfnamefont {J.}~\bibnamefont {Bekaert}}, \bibinfo {author} {\bibfnamefont {A.}~\bibnamefont {\ifmmode \check{S}\else \v{S}\fi{}olaji\ifmmode~\acute{c}\else \'{c}\fi{}}}, \bibinfo {author} {\bibfnamefont {J.}~\bibnamefont {Pe\ifmmode \check{s}\else \v{s}\fi{}i\ifmmode~\acute{c}\else \'{c}\fi{}}}, \bibinfo {author} {\bibfnamefont {A.}~\bibnamefont {Wang}}, \bibinfo {author} {\bibfnamefont {M.}~\bibnamefont {\ifmmode \check{S}\else \v{S}\fi{}\ifmmode \acute{c}\else \'{c}\fi{}epanovi\ifmmode~\acute{c}\else \'{c}\fi{}}}, \bibinfo {author} {\bibfnamefont {A.~M.~M.}\ \bibnamefont {Abeykoon}},
  \bibinfo {author} {\bibfnamefont {M.~V.}\ \bibnamefont {Milo\ifmmode \check{s}\else \v{s}\fi{}evi\ifmmode~\acute{c}\else \'{c}\fi{}}}, \bibinfo {author} {\bibfnamefont {C.}~\bibnamefont {Petrovic}}, \bibinfo {author} {\bibfnamefont {Z.~V.}\ \bibnamefont {Popovi\ifmmode~\acute{c}\else \'{c}\fi{}}},\ and\ \bibinfo {author} {\bibfnamefont {R.}~\bibnamefont {Hackl}},\ }\bibfield  {title} {\bibinfo {title} {Evolution of lattice, spin, and charge properties across the phase diagram of ${\mathrm{fese}}_{1\ensuremath{-}x}{\mathrm{s}}_{x}$},\ }\href {https://doi.org/10.1103/PhysRevB.106.094510} {\bibfield  {journal} {\bibinfo  {journal} {Phys. Rev. B}\ }\textbf {\bibinfo {volume} {106}},\ \bibinfo {pages} {094510} (\bibinfo {year} {2022})}\BibitemShut {NoStop}%
\bibitem [{\citenamefont {Abramovitch}\ \emph {et~al.}(2024)\citenamefont {Abramovitch}, \citenamefont {Mravlje}, \citenamefont {Zhou}, \citenamefont {Georges},\ and\ \citenamefont {Bernardi}}]{EPC_Georges1}%
  \BibitemOpen
  \bibfield  {author} {\bibinfo {author} {\bibfnamefont {D.~J.}\ \bibnamefont {Abramovitch}}, \bibinfo {author} {\bibfnamefont {J.}~\bibnamefont {Mravlje}}, \bibinfo {author} {\bibfnamefont {J.-J.}\ \bibnamefont {Zhou}}, \bibinfo {author} {\bibfnamefont {A.}~\bibnamefont {Georges}},\ and\ \bibinfo {author} {\bibfnamefont {M.}~\bibnamefont {Bernardi}},\ }\bibfield  {title} {\bibinfo {title} {Respective roles of electron-phonon and electron-electron interactions in the transport and quasiparticle properties of ${\mathrm{srvo}}_{3}$},\ }\href {https://doi.org/10.1103/PhysRevLett.133.186501} {\bibfield  {journal} {\bibinfo  {journal} {Phys. Rev. Lett.}\ }\textbf {\bibinfo {volume} {133}},\ \bibinfo {pages} {186501} (\bibinfo {year} {2024})}\BibitemShut {NoStop}%
\bibitem [{\citenamefont {Lee}(2015)}]{Lee:2015}%
  \BibitemOpen
  \bibfield  {author} {\bibinfo {author} {\bibfnamefont {D.-H.}\ \bibnamefont {Lee}},\ }\bibfield  {title} {\bibinfo {title} {What makes the tc of fese/srtio3 so high?},\ }\href {https://doi.org/10.1088/1674-1056/24/11/117405} {\bibfield  {journal} {\bibinfo  {journal} {Chinese Physics B}\ }\textbf {\bibinfo {volume} {24}},\ \bibinfo {eid} {117405} (\bibinfo {year} {2015})}\BibitemShut {NoStop}%
\bibitem [{\citenamefont {Zhang}\ \emph {et~al.}(2017)\citenamefont {Zhang}, \citenamefont {Liu}, \citenamefont {Chen}, \citenamefont {Xie}, \citenamefont {He}, \citenamefont {Tang}, \citenamefont {He}, \citenamefont {Li}, \citenamefont {Jia}, \citenamefont {Rebec}, \citenamefont {Ma}, \citenamefont {Yan}, \citenamefont {Hashimoto}, \citenamefont {Lu}, \citenamefont {Mo}, \citenamefont {Hikita}, \citenamefont {Moore}, \citenamefont {Hwang}, \citenamefont {Lee},\ and\ \citenamefont {Shen}}]{FeSe-STO-natcom}%
  \BibitemOpen
  \bibfield  {author} {\bibinfo {author} {\bibfnamefont {C.}~\bibnamefont {Zhang}}, \bibinfo {author} {\bibfnamefont {Z.}~\bibnamefont {Liu}}, \bibinfo {author} {\bibfnamefont {Z.}~\bibnamefont {Chen}}, \bibinfo {author} {\bibfnamefont {Y.}~\bibnamefont {Xie}}, \bibinfo {author} {\bibfnamefont {R.}~\bibnamefont {He}}, \bibinfo {author} {\bibfnamefont {S.}~\bibnamefont {Tang}}, \bibinfo {author} {\bibfnamefont {J.}~\bibnamefont {He}}, \bibinfo {author} {\bibfnamefont {W.}~\bibnamefont {Li}}, \bibinfo {author} {\bibfnamefont {T.}~\bibnamefont {Jia}}, \bibinfo {author} {\bibfnamefont {S.~N.}\ \bibnamefont {Rebec}}, \bibinfo {author} {\bibfnamefont {E.~Y.}\ \bibnamefont {Ma}}, \bibinfo {author} {\bibfnamefont {H.}~\bibnamefont {Yan}}, \bibinfo {author} {\bibfnamefont {M.}~\bibnamefont {Hashimoto}}, \bibinfo {author} {\bibfnamefont {D.}~\bibnamefont {Lu}}, \bibinfo {author} {\bibfnamefont {S.-K.}\ \bibnamefont {Mo}}, \bibinfo {author} {\bibfnamefont {Y.}~\bibnamefont {Hikita}}, \bibinfo {author} {\bibfnamefont
  {R.~G.}\ \bibnamefont {Moore}}, \bibinfo {author} {\bibfnamefont {H.~Y.}\ \bibnamefont {Hwang}}, \bibinfo {author} {\bibfnamefont {D.}~\bibnamefont {Lee}},\ and\ \bibinfo {author} {\bibfnamefont {Z.}~\bibnamefont {Shen}},\ }\bibfield  {title} {\bibinfo {title} {Ubiquitous strong electron--phonon coupling at the interface of fese/srtio3},\ }\href {https://doi.org/10.1038/ncomms14468} {\bibfield  {journal} {\bibinfo  {journal} {Nature Communications}\ }\textbf {\bibinfo {volume} {8}},\ \bibinfo {pages} {14468} (\bibinfo {year} {2017})}\BibitemShut {NoStop}%
\bibitem [{\citenamefont {Moghadas}\ \emph {et~al.}(2025)\citenamefont {Moghadas}, \citenamefont {Reitner}, \citenamefont {Wehling}, \citenamefont {Sangiovanni}, \citenamefont {Ciuchi},\ and\ \citenamefont {Toschi}}]{arxiv-EPC}%
  \BibitemOpen
  \bibfield  {author} {\bibinfo {author} {\bibfnamefont {E.}~\bibnamefont {Moghadas}}, \bibinfo {author} {\bibfnamefont {M.}~\bibnamefont {Reitner}}, \bibinfo {author} {\bibfnamefont {T.}~\bibnamefont {Wehling}}, \bibinfo {author} {\bibfnamefont {G.}~\bibnamefont {Sangiovanni}}, \bibinfo {author} {\bibfnamefont {S.}~\bibnamefont {Ciuchi}},\ and\ \bibinfo {author} {\bibfnamefont {S.}~\bibnamefont {Toschi}},\ }\bibfield  {title} {\bibinfo {title} {Effective enhancement of the electron-phonon coupling driven by nonperturbative electronic density fluctuations},\ }\href {https://arxiv.org/abs/2503.12113} {\bibfield  {journal} {\bibinfo  {journal} {arXiv:2503.12113}\ } (\bibinfo {year} {2025})}\BibitemShut {NoStop}%
\bibitem [{\citenamefont {Lee}\ \emph {et~al.}(2014{\natexlab{b}})\citenamefont {Lee}, \citenamefont {Schmitt}, \citenamefont {Moore}, \citenamefont {Johnston}, \citenamefont {Cui}, \citenamefont {Li}, \citenamefont {Yi}, \citenamefont {Liu}, \citenamefont {Hashimoto}, \citenamefont {Zhang}, \citenamefont {Lu}, \citenamefont {Devereaux}, \citenamefont {Lee},\ and\ \citenamefont {Shen}}]{FeSC-EPC1}%
  \BibitemOpen
  \bibfield  {author} {\bibinfo {author} {\bibfnamefont {J.~J.}\ \bibnamefont {Lee}}, \bibinfo {author} {\bibfnamefont {F.~T.}\ \bibnamefont {Schmitt}}, \bibinfo {author} {\bibfnamefont {R.~G.}\ \bibnamefont {Moore}}, \bibinfo {author} {\bibfnamefont {S.}~\bibnamefont {Johnston}}, \bibinfo {author} {\bibfnamefont {Y.~T.}\ \bibnamefont {Cui}}, \bibinfo {author} {\bibfnamefont {W.}~\bibnamefont {Li}}, \bibinfo {author} {\bibfnamefont {M.}~\bibnamefont {Yi}}, \bibinfo {author} {\bibfnamefont {Z.~K.}\ \bibnamefont {Liu}}, \bibinfo {author} {\bibfnamefont {M.}~\bibnamefont {Hashimoto}}, \bibinfo {author} {\bibfnamefont {Y.}~\bibnamefont {Zhang}}, \bibinfo {author} {\bibfnamefont {D.~H.}\ \bibnamefont {Lu}}, \bibinfo {author} {\bibfnamefont {T.~P.}\ \bibnamefont {Devereaux}}, \bibinfo {author} {\bibfnamefont {D.~H.}\ \bibnamefont {Lee}},\ and\ \bibinfo {author} {\bibfnamefont {Z.~X.}\ \bibnamefont {Shen}},\ }\bibfield  {title} {\bibinfo {title} {Interfacial mode coupling as the origin of the enhancement of tc in
  fese films on srtio3},\ }\href {https://doi.org/10.1038/nature13894} {\bibfield  {journal} {\bibinfo  {journal} {Nature}\ }\textbf {\bibinfo {volume} {515}},\ \bibinfo {pages} {245} (\bibinfo {year} {2014}{\natexlab{b}})}\BibitemShut {NoStop}%
\end{thebibliography}%

\section{End Matter}

\begin{figure*}[t]
\centering
\includegraphics[width=1\textwidth, angle=0]{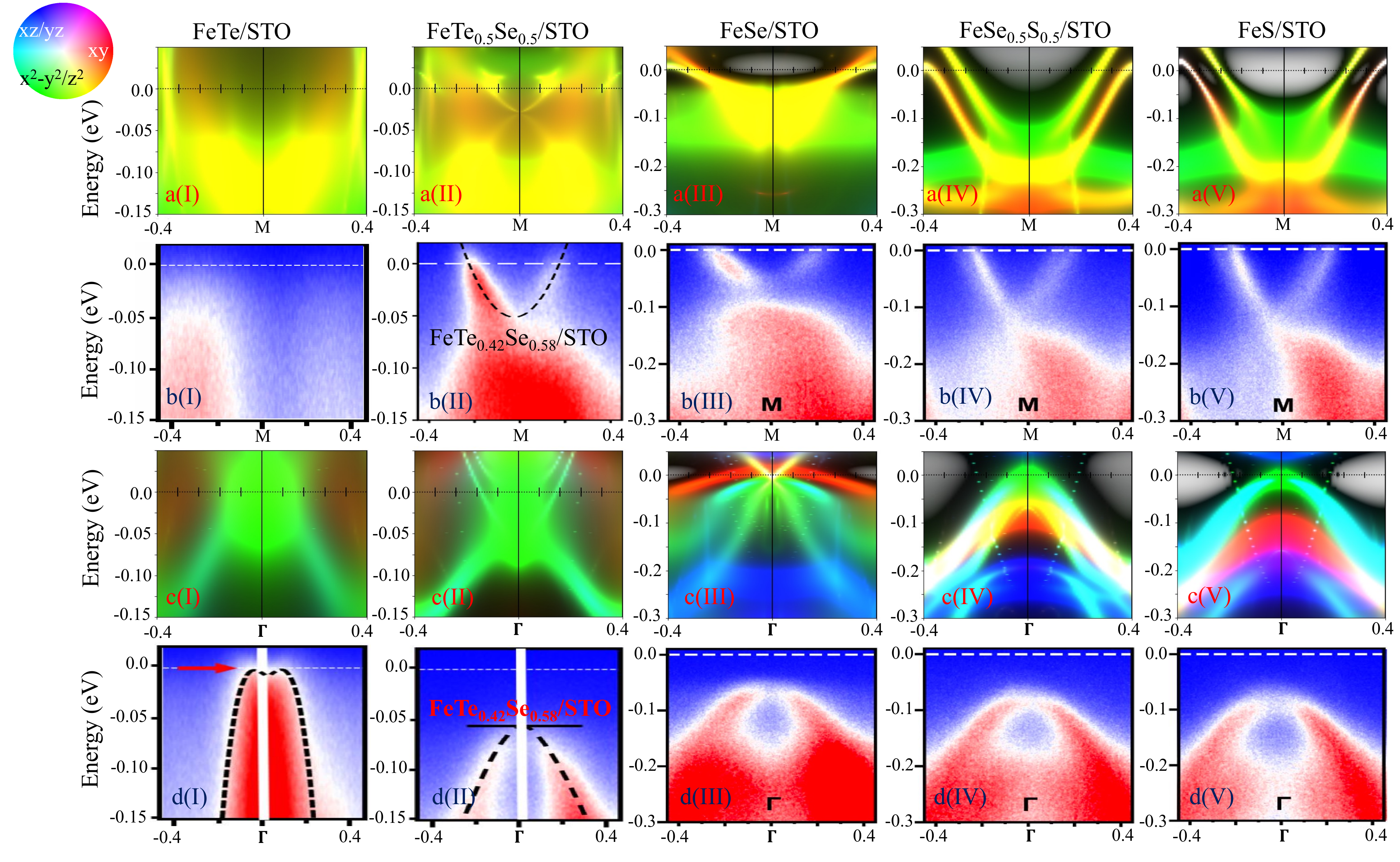} 
\caption{Systematic comparison of spectral functions obtained from eDMFT computations [a, c(I--V)] and APRES [b, d(I--V)] as a function of S/Te-alloying in single layer FeX/STO.}\label{fig4_em}
\end{figure*}

{\it Spectral Function Evolution with S/Te alloying: }In Fig.~\ref{fig4_em}, we illustrate this evolution, showing both experimental results from ARPES and theoretical predictions from eDMFT calculations. The eDMFT approach successfully captures key trends in the degree of electronic correlation and charge transfer associated with S/Te substitution, in good agreement with ARPES observations.

With increasing Te content, we observe a clear enhancement in electronic correlations, reflected in both the electron (Figs.~\ref{fig4_em}b(I)--b(V)) and hole (Figs.~\ref{fig4_em}d(I)--d(V)) pockets. This manifests as broader spectral features in the SF, indicative of increased quasiparticle scattering and reduced coherence, especially in comparison to the sharper and more well-defined bands seen in pure FeSe/STO. These changes are consistent with the mass enhancement trends shown in Fig.~\ref{fig4}\textbf{b}, where Te substitution leads to heavier effective masses. For S alloying, the opposite trend is observed: spectral features become increasingly coherent, suggesting a suppression of many-body scattering effects as the system moves away from strong correlations. Notably, the eDMFT-predicted SFs capture this modulation of correlation strength and its impact on band coherence with impressive accuracy. One exception occurs in the region near the $\Gamma$-point for the hole pockets in FeSe/STO, FeSe$_{0.5}$S$_{0.5}$/STO, and FeS/STO (Figs.~\ref{fig4_em}c(III)--c(V)), where slight discrepancies between theory and experiment are observed. These deviations may be attributed to subtleties in the substrate-induced band renormalization or disorder effects in the alloyed samples, which are not fully captured in the current theoretical model.

The overall agreement between eDMFT and ARPES is excellent, particularly around the $M$-point (Figs.~\ref{fig4_em}a(I)--a(V)) where the electron pockets reside. The SFs computed for FeTe/STO and FeTe$_{0.5}$Se$_{0.5}$/STO around the $\Gamma$-point (Figs.~\ref{fig4_em}c(I) and c(II)) also show good correspondence with the ARPES data (Figs.~\ref{fig4_em}d(I) and d(II)), reinforcing the validity of the eDMFT framework in describing the correlated electronic structure across this family of superconducting heterostructures.

\end{document}